\documentclass[aps,prd,floatfix,nofootinbib,superscriptaddress,twocolumn,showkeys]{revtex4-2}

\usepackage{amsmath}
\usepackage{listings}
\usepackage{amssymb}
\allowdisplaybreaks
\usepackage{float}
\usepackage{makeidx}
\usepackage{amsfonts}
\usepackage[usenames,dvipsnames]{pstricks}
\usepackage{subfigure}
\usepackage{epsfig}
\usepackage{pst-grad} 
\usepackage{mathtools}
\usepackage[colorlinks,hyperindex]{hyperref}
\usepackage{array}
\usepackage{color}

\hypersetup
{	colorlinks,%
citecolor=black,%
linkcolor=black,%
urlcolor=black,%
}

\usepackage{allpurpose}

\usepackage{flushend}

\usepackage[utf8]{inputenc}

\newcommand\nn{{\nonumber}}
\newcommand\E{{E}}
\newcommand\te{\mathrm{e}}
\newcommand\hata{\hat{a}}
\newcommand\mad{\mathrm{d}}
\newcommand\mas{\mathrm{s}}

\def\XXint#1#2#3{{\setbox0=\hbox{$#1{#2#3}{\int}$}	\vcenter{\hbox{$#2#3$}}\kern-.5\wd0}}

\usepackage{listings}

\usepackage{ulem}
\usepackage{color}

\newcommand{\delred}[1]{{\color{red}{\ifmmode\text{\sout{\ensuremath{#1}}}\else\sout{#1}\fi}}}

\begin{document}

\title{Off-equatorial deflections and gravitational lensing. I. In Kerr spacetime and effect of spin}

\author{Tingyuan Jiang}
\thanks{These authors contributed equally to this work.}
\address{School of Physics and Technology, Wuhan University, Wuhan, 430072, China}

\author{Xiaoge Xu}
\thanks{These authors contributed equally to this work.}
\address{School of Physics and Technology, Wuhan University, Wuhan, 430072, China}

\author{Junji Jia}
\email[Corresponding author:~]{junjijia@whu.edu.cn}
\address{Department of Astronomy \& MOE Key Laboratory of Artificial Micro- and Nano-structures, School of Physics and Technology, Wuhan University, Wuhan, 430072, China}

\date{\today}

\begin{abstract}
	This paper investigates off-equatorial plane deflections and gravitational lensing of both null signals and massive particles in Kerr spacetime in the weak deflection limit, with the finite distance effect of the source and detector taken into account. This is the effect caused by the fact that both the source and detector are located at finite distances from the lens, while many researchers often use the deflection angle for infinite distances from sources and detectors. The deflection in both the $\phi$ and $\theta$ directions is computed as power series of $M/r_0$ and $r_0/r_{\mathrm{s,d}}$, where $M,\,r_{\mathrm{s,d}}$ are the spacetime mass and source and detector radii respectively, and $r_0$ is the minimal radial coordinate of the trajectory. The coefficients of these series are simple trigonometric functions of $\theta_\te$, the extreme value of the $\theta$ coordinate of the trajectory. A set of exact gravitational lensing equations is used to solve for $r_0$ and $\theta_\te$ for given deviation angles $\delta\theta$ and $\delta\phi$ of the source, and two lensed images are always obtained. The apparent angles and their magnifications of these images, and the time delays between them are solved and their dependence on various parameters, especially spacetime spin $\hat{a}$ are analyzed in great detail. It is found that there generally exist two critical spacetime spin values that separate the case of test particles reaching the detector from different sides of the $z$ axis from the cases in which the images appear from the same side in the celestial plane. Three potential applications of these results are discussed.
\end{abstract}


\maketitle

\section{Introduction}

Deflection and gravitational lensing (GL) of lightrays are fundamental features of signals' motion in curved spacetimes. The early confirmation of the former established general relativity as a more accurate description of gravity \cite{Dyson:1920cwa}. The GL has become an important tool in astronomy, from measuring the lens mass  
and mass distribution 
\cite{Bartelmann:1999yn}
to study the properties of dark mass and dark energy 
\cite{Metcalf:2001ap, Hoekstra:2008db}. With advancement of the observation technologies, especially the fast development of black hole (BH) observations (e.g. Event Horizon Telescope, see
\cite{EventHorizonTelescope:2019dse,EventHorizonTelescope:2022wkp}), the physics of bending and GL of test particles become essential for the correct understanding of the relevant observation results. 

Theoretically, the deflection and GL of test particles are most easily understood and thoroughly studied in static and spherically symmetric (SSS) spacetimes, or in the equatorial plane of stationary and axisymmetric spacetimes. Different techniques, such as the Gauss-Bonnet theorem-based geometrical method \cite{GW2008,Werner2012,Li:2019vhp,Crisnejo:2019ril,Li:2020dln} and perturbative methods \cite{Beachley:2018tux, Jia:2020xbc,Liu:2020mkf}, have been used to study deflections of not only null signals but maissive particles. Various effects, such as the finite distance effect of the source and detectors \cite{Ishihara:2016vdc,Li:2019qyb,Huang:2020trl}, and effects of the spacetime parameters (spin, charges \cite{Pang:2018jpm}, magnetic field \cite{Li:2022cpu} etc.) and properties of test particle  (spin \cite{Zhang:2022rnn} and charge \cite{Li:2020ozr,Xu:2021rld}), have been extensively investigated in recent years. Observationally, however, Kerr BH is still one of the most simple and natural BH candidates considered by astronomers \cite{EventHorizonTelescope:2019dse, EventHorizonTelescope:2022wkp}. 

The deflection of test particle and GL of light rays in Kerr spacetime have also been intensively studied. Although relevant numerical packages can often yield the general motions in this spacetime \cite{Yang:2013yoa}, analytical works about the deflection and GL are more concentrated on the motion in the (quasi-)equatorial plane \cite{Iyer:2009hq, Aazami:2011tu, Aazami:2011tw, He:2016xiu, Renzini:2017bqg, Liu:2017fjx, Beachley:2018tux, Hsiao:2019ohy}. 
For analytical works dealing with non-equatorial plane motions in the Kerr spacetime, 
Wilkins investigated the frequencies of the bound orbit in both the $\theta$ and $\phi$ directions \cite{Wilkins:1972rs}.  
Fujita and Hikida studied the bound timelike orbit solution in terms of Mino time \cite{Fujita:2009bp}. Hackmann and Xu classified motions of test particles in Kerr and KN spacetimes \cite{Hackmann:2010tqa, Hackmann:2013pva}.  For works most relevant to the present paper, Bray calculated the deflection angles of null rays in Kerr spacetime in terms of the conserved constant of the motion \cite{Bray:1985ew}. Sereno and De Luca updated this work and solve the image positions using some approximate geometrical relations \cite{Sereno:2006ss}. Kraniotis found image locations for observers with certain particular latitudes \cite{Kraniotis:2010gx}. Gralla and Lupsasca discussed the highly-bent rays and properties of photon rings in Kerr BH spacetime \cite{Gralla:2019drh}.

The above works however did not systematically take into account the finite distance effect of the source and observer. The lens equations used were also based on the first-order approximation of the geometrical relations linking the angle of the source against the lens-detector axis and the apparent angle. In this work, we will develop a perturbative method that can compute the deflection and GL of test particles with arbitrary orientation directions in the Kerr spacetime. Moreover, the method can take into account the finite distance effect of the source and observer, which allows the use of the exact GL equations to obtain the image positions, their magnifications and time delays. Furthermore, we will not limit the trajectories to light rays but deal with test particles with general asymptotic velocity, i.e., massive particles are also included. 
The deflection and GL of massive particles have drawn more attention in recent years \cite{Glicenstein:2017lrm,Pang:2018jpm, Li:2019qyb,Li:2019vhp,Crisnejo:2019ril,Li:2020dln,Jia:2020xbc, Liu:2020mkf} with the fast development of neutrino \cite{IceCube:2018dnn,IceCube:2018cha, IceCube:2022der} and cosmic ray (see \cite{AMS:2021nhj} and references therein)
observation technologies, and the discovery of gravitational waves and the fact that gravitational waves in some beyond general relativity theories are massive \cite{LIGOScientific:2016aoc, LIGOScientific:2017vwq}.  

In this work, we will show that the deflection angles in Kerr spacetime with mass $M$ in the weak deflection limit (WDL) in both the $\phi$ and $\theta$ directions can be expressed in quasi-power series forms of $M/r_0$ and $r_0/r_{\mathrm{s,d}}$ where $r_0,\,r_{\mathrm{s,d}}$ are the minimal radial coordinate of the trajectory and the source and detector radial coordinates respectively. The coefficients of these series are functions of $\hat{a}$, the spin angular momentum per unit mass, and $v$ (or $E$), the asymptotic velocity of the test particles . After solving a set of exact GL equations, we will determine the image positions and magnifications of a source located at arbitrary azimuthal and zenith angles using these deflections. The dependence of these quantities and the time delay between images on the spacetime spin size and its orientation and other parameters will be shown explicitly. We also used these results to discuss some potential applications in astronomical observations.

The work is organized as follows. In Sec. \ref{sec:Preliminaries} we introduce the basic setup of the problem. In Sec. \ref{sec:pertmeth} the perturbative method is developed and used to express the deflections as power series. The GL equations are solved in Sec. \ref{sec:gltd} to obtain the image locations, magnifications and time delays. The effects of various parameters on them are also investigated. Sec. \ref{sec:appl} discusses a few potential applications of the results and concludes the paper. Throughout the work, we use the natural units $G=c=1$.

\section{Preliminaries \label{sec:Preliminaries}}

The Kerr spacetime with the  Boyer-Lindquist coordinates $(t,r,\theta,\phi)$ can be described by the following metric 
\begin{align}
	\dd s^2=&-\frac{\Delta}{\Sigma}\left(\dd t-a\sin^2\theta \dd\phi\right)^2+\frac{\Sigma}{\Delta}\dd r^2+\Sigma \dd\theta^2\nonumber\\
	& +\frac{\sin^2\theta}{\Sigma}\left[\left(r^2+a^2\right)\dd\phi-a\dd t\right]^2,\label{eq:kerrmetric}
\end{align}
where 
\begin{align}
	&\Delta(r)=r^2-2Mr+a^2,\\
	&\Sigma(r,\theta)=r^2+a^2\cos^2\theta
\end{align}
and $a=J/M$ is the angular momentum per unit mass of the BH, with $M$ being its total mass. 
The motion of test particles in this spacetime is governed by the geodesic equation 
\begin{equation}
	\frac{\dd^2 x^\rho}{\dd\sigma^2} + \Gamma_{\mu,\nu}^{\rho}\frac{\dd x^{\mu}}{\dd\sigma}\frac{d x^{\nu}}{d\sigma} = 0, \label{eq:geodesticdef}
\end{equation}
where $\sigma$ is the proper time of massive particles or affine parameter of null signals. Using the metric \eqref{eq:kerrmetric}, this becomes after the first integrals 
\cite{Kapec:2019hro}
\begin{subequations}
	\begin{align}
		&\Sigma^2\left(  \frac{\dd r}{\dd\sigma}\right)^2 = R(r), \label{eq:rgeodesic}\\
		&\Sigma^2\left(  \frac{\dd\cos\theta}{\dd\sigma}\right)^2 =\Theta(\cos\theta),\label{eq:thetageodesic}\\
		&\Sigma \frac{\dd\phi}{\dd\sigma}=\frac{2aMrE-a^2L}{\Delta}+L\csc^2\theta,\label{eq:phigeodesic}\\
		&\Sigma \frac{\dd t}{\dd\sigma}=\frac{E(r^2+a^2)^2-2aLMr}{\Delta}-Ea^2 \sin^2\theta,\label{eq:tgeodesic}
	\end{align}
\end{subequations}
where 
\begin{align}		
	R(r)=&\left[E\left(r^2+a^2\right)-aL\right]^2-\Delta\left(K+m^2 r^2\right), \label{eq:Rdef}\\
	\Theta(\cos\theta)=&\left(1-\cos^2\theta\right)\left[K-a^2m^2\cos^2\theta+2LaE\right.\nonumber\\
	&\left.-a^2E^2\left(1-\cos^2\theta\right)\right] -L^2\label{eq:Thetadef}
\end{align}
and $m,\,E,\,L,\,K$ are respectively the mass, the conserved energy and angular momentum of the test particle, and the Carter constant. 
In asymptotically Minkowski spacetimes including the Kerr one, $E$ can be related to the asymptotic velocity $v$ (the spatial components of the four-velocity) of the massive particle observed by a static observer far from the center, using the relation
\begin{align}
	E=\frac{m}{\sqrt{1-v^2}}.
\end{align}
While for the null signal, $m$ approaches zero but $v$ approaches 1, and $E$ is still finite. For the equations and results throughout this paper, we can always obtain the null limit by taking $v\to 1$. 
One of the main motivations for the work is to obtain the deflection of the test particles that are not restricted to the equatorial plane. For this purpose, using Eqs. \eqref{eq:rgeodesic} and \eqref{eq:thetageodesic}, we first get 
\begin{equation}
	\frac{s_{r}\dd r}{\sqrt{R(r)}}= \frac{s_{\theta}\dd \cos\theta}{\sqrt{\Theta(\cos\theta)}}.   \label{eq:difftheta}
\end{equation}
Here, the $s_\theta$ and $s_r$ are two signs introduced when taking the square root in Eqs. \eqref{eq:rgeodesic} and \eqref{eq:thetageodesic} respectively.
Using Eqs. \eqref{eq:rgeodesic} and \eqref{eq:difftheta} in the first and last terms respectively on the right-hand side of Eq. \eqref{eq:phigeodesic}, we obtain 
\begin{align}
	\dd \phi =& \frac{2aMrE-a^2L}{\Delta} \frac{s_{r}\dd r}{\sqrt{R(r)}} +\frac{L}{1-\cos^2\theta}\frac{s_{\theta}\dd \cos\theta}{\sqrt{\Theta(\cos\theta)}}. \label{eq:diffphi}
\end{align} 
When proper initial conditions are given, integrating Eqs. \eqref{eq:difftheta} and \eqref{eq:diffphi} from the source to the detector will yield the deflection in the $\phi$ and $\theta$ directions respectively. If we let $r_\mas$ (or $r_\mad$) vary, then knowing the integral results of Eqs. \eqref{eq:difftheta} and \eqref{eq:diffphi} is equivalent to knowing the solutions $\phi(r)$ and $\theta(r)$. 

Before carrying out more detailed computations, there are a few comments regarding these equations and their integrals.
The first is to note that according to Ref. \cite{Hackmann:2010tqa}, there are quite a few kinds of motion in the Kerr spacetime when the particle is not limited to the equatorial plane. The kind we will study is classified as the IVb case, which is a flyby orbit. In other words, the particle will come from a large distance to reach a periapsis and then return to another large distance. In this case, we can adjust the orbit parameters so that the periapsis is far from the event horizon and therefore the deflection of the test particle is generally weak, which is essential for the feasibility of the weak field perturbative study. In this limit, we can safely assume that the $\theta$ coordinate will only experience one local extremum along the entire trajectory. We denote this extreme value as $\theta_\te$. If the signal flies by the lens from above (or below) the equatorial plane, $\theta_\te$ will be a minimum (or maximum), while $\cos\theta_\te$ will be a local maximum (or minimum). 

With the above consideration, we can integrate Eqs. \eqref{eq:difftheta} and \eqref{eq:diffphi} from the source located at $(r_\mas,\,\theta_\mas,\,\phi_\mas)$ to the detector at $(r_\mad,\,\theta_\mad,\,\phi_\mad)$ to obtain the following relation 
\begin{equation}
	\left(\int_{r_0}^{r_\mas} + \int_{r_0}^{r_\mad}\right) \frac{\dd r}{\sqrt{R(r)}} = \left(\int_{\theta_\mas}^{\theta_\te} + \int_{\theta_\mad}^{\theta_\te}\right)\frac{s_{r\theta}\dd \cos\theta}{\sqrt{\Theta(\cos\theta)}},
	\label{eq:thetaintdef}
\end{equation}
\begin{align}
	&s_l\int_{\phi_\mas}^{\phi_\mad}\dd \phi = \left(\int_{r_0}^{r_\mas} + \int_{r_0}^{r_\mad}\right) \frac{2aMrE-a^2L}{\Delta\sqrt{R(r)}}  \dd r \nonumber\\
	& + \left(\int_{\cos\theta_\mas}^{\cos\theta_\te} + \int_{\cos\theta_\mad}^{\cos\theta_\te}\right)\frac{s_{r\theta}L}{1-\cos^2\theta}\frac{\dd \cos\theta}{\sqrt{\Theta(\cos\theta)}}.
	\label{eq:phiintdef}
\end{align}
Here $r_0$ is the minimum radial coordinate along the trajectory, and $\theta_\te$ is the extreme value of the $\theta$ coordinate. $s_{r\theta}=\pm1$ and $s_l=\pm1$ are the signs induced from Eqs. \eqref{eq:difftheta} and \eqref{eq:diffphi} when carrying out the integrals. Here $s_l$ is the same as the sign of the orbital angular momentum introduced in Eq. \eqref{eq:ltor0}. $r_0$ and $\theta_\te$ can be related to the conserved constants $L,\,K$ and $E$ through their definitions 
\begin{align}
	\left.\frac{\dd r}{\dd\sigma}\right\vert_{r=r_0} = 0, ~~
	\left.\frac{\dd\cos\theta}{\dd\sigma}\right\vert_{\theta=\theta_{\mathrm e }} = 0.\label{eq:r0thetamdef}
\end{align}
From Eqs. \eqref{eq:rgeodesic} and \eqref{eq:thetageodesic}, we see that the above equation is equivalent to finding the roots of the right-hand sides of Eqs. \eqref{eq:Rdef} and \eqref{eq:Thetadef} by setting them equal to zero. Using these, we have obtained the analytical expression for $\theta_\te$ in terms of $L,\,K$ and $E$ 
\begin{align}
	&\cos^2\theta_e = \frac{1}{2 a^2 \left(E^2-m^2\right)}\Big[a^2 \left(2 E^2-m^2\right)-2 aE L-K \nn\\
	&+\lcb (a^2 m^2-K)^2+4aL\lsb EK-a(aE-L)m^2\rsb \rcb^{1/2}\Big].
\end{align}
$r_0$ is a root of and order four polynomial and too lengthy to show here. Note that  $L$ and $K$ can be expressed in terms of $r_0,\,\theta_\te$ and $E$ 
\begin{align}
	L=&\frac{ s_l\sin\theta_{\mathrm e }\chi-2 a E \sin^2\theta_{\mathrm e } M r_0 }{\Sigma(r_0,\theta_{\mathrm e })-2 M r_0}, \label{eq:ltor0}\\
	K
	=&a^2m^2\cos^2\theta_{\mathrm e }+(L \csc \theta_{\mathrm e } -a E \sin \theta_{\mathrm e } )^2,
	\label{eq:ktor0}
\end{align}
where
\begin{align}
	\chi=\sqrt{\Sigma(r_0,\theta_{\mathrm e })\Delta(r_0)\lsb\Sigma(r_0,\theta_{\mathrm e })(E^2-m^2)+2M m^2 r_0\rsb}\nn
\end{align}
and the $s_l$ in front of $\chi$ is valid in the WDL. These relations can be used to replace $L$ and $K$ in integrals \eqref{eq:thetaintdef} and \eqref{eq:phiintdef} later.

Among the six coordinates $(r_\mas,\theta_\mas,\phi_\mas)$ and $(r_\mad,\theta_\mad,\phi_\mad)$, we will assume that $r_\mas,\,r_\mad$ and $\theta_\mas$ are known. The $\phi_\mas$ and $\phi_\mad$ are unnecessary to know {\it a priori} and indeed $\Delta\phi\equiv\phi_\mad-\phi_\mas$ is the deflection angle we desire to solve. We also note that the integral in Eq. \eqref{eq:phiintdef} is exactly the deflection $\Delta \phi$ in the $\phi$ direction. For the deflection $\Delta\theta\equiv \theta_\mad+\theta_\mas-\pi$ in the $\theta$ direction, we see from Eq. \eqref{eq:thetaintdef} that once $r_\mas,\,r_\mad$ and $\theta_\mas$ are given, and if one can carry out the integral in this equation, then solving the resultant algebraic equation will allow us to determine $\theta_\mad$ and consequently $\Delta\theta$. 

One of the main efforts of this work is to find proper, tractable ways to carry out these integrals. We will show in the next section that there exists a perturbative method to systematically approximate these deflections, and the result takes a dual series form of $M/r_0$ and $r_0/r_{\mathrm{s,d}}$. 

\section{Perturbative method and results\label{sec:pertmeth}}

\subsection{Perturbative expansion method}

The key to successfully carrying out integrals \eqref{eq:thetaintdef} and \eqref{eq:phiintdef} is to find a proper way to expand the integrands into integrable series which allows approximations to any desired accuracy. 
The WDL provides a natural expansion parameter, the ratio $M/r_0$. Before carrying out this expansion, let us also point out that the orbital angular momentum $L$ and the Cater constant $K$ of the test particle are not easily measurable. Therefore one has to replace them in Eqs. \eqref{eq:thetaintdef} and \eqref{eq:phiintdef} by Eqs. \eqref{eq:ltor0} and \eqref{eq:ktor0}. For simpler notations, introducing the new integration variables, 
\begin{align}
    p\equiv r_0/r,~c \equiv \cos\theta,
\end{align} as well as the auxiliary notations 
\begin{align}
	p_\mathrm{s,d} &= r_0/r_{\mathrm{s,d}},~c_\mathrm{s,d,e} = \cos\theta_\mathrm{s,d,e},\nn\\
	~s_\mathrm{s,d,e} &= \sin\theta_\mathrm{s,d,e},~t_\mathrm{s,d,e} = \tan\theta_\mathrm{s,d,e} ,\label{eq:auxdef}
\end{align}
and then carrying out the expansions using $M/r_0$ as a small parameter, Eqs. \eqref{eq:thetaintdef} and \eqref{eq:phiintdef} become
\begin{align}
	&\left(\int_{1}^{p_\mas} + \int_{1}^{p_\mathrm{d}}\right)\sum_{i = 1}^{\infty}\frac{f_{r,i}(p)}{ (1 + p)^{i-1} \sqrt{1-p^2}  } \left(\frac{M}{r_0}\right)^i \dd p\nonumber\\
	&=  \left(\int_{c_\mas}^{c_\te} + \int_{c_\mathrm{d}}^{c_\te}\right)\sum_{i = 1}^{\infty} \frac{s_{r\theta}f_{\theta,i}(c)}{ \sqrt{c_\te^2-c^2}} \left(\frac{M}{r_0}\right)^i \dd c,
	\label{eq:seriesdifftheta}
\end{align}
\begin{align}
	\Delta \phi =& \left(\int_{1}^{p_\mas} + \int_{1}^{p_\mathrm{d}}\right) \sum_{i = 2}^{\infty} \frac{g_{r,i}(p)}{(1 + p)^{i-2}\sqrt{1-p^2}} \left(\frac{M}{r_0}\right)^i \dd p \nonumber\\
	&+ \left(\int_{c_\mas}^{c_\te} + \int_{c_\mathrm{d}}^{c_\te}\right) \sum_{i = 0}^{\infty}  \frac{s_{r\theta}g_{\theta,i}(c) s_\te}{\sqrt{c_\te^2-c^2}}\left(\frac{M}{r_0}\right)^i \dd c,
	\label{eq:seriesdiffphi}
\end{align}
where $f_{r,i},\,f_{\theta,i},\,g_{r,i},\,g_{\theta,i}$ are the Taylor expansion coefficients of the integrands whose exact forms can be worked out easily. Here we list the first few orders of them
\begin{subequations}
	\begin{align}
		f_{r,1} =& \frac{1}{M v E}, ~f_{r,2} = \frac{p[1-(1+p)v^2]}{M v^3 E}, ~\cdots,\\
		f_{\theta,1} =& \frac{1}{M v E},~ f_{\theta,2} = -\frac{1}{M v^3 E}, ~\cdots,\\
		g_{r,2} =& \hat{a} p \left(-\frac{2 s_l}{v}+\hat{a}s_\te  p\right),~\cdots,\\
		g_{\theta,0} =& \frac{1}{1-c^2},~ g_{\theta,1} = 0,~g_{\theta,2} = \frac{\hat{a}^2}{2}, ~\cdots,
	\end{align}
\end{subequations}
where $\hat{a} \equiv a/M$.

Since all $f_{r,i}$ and $g_{r,i}$ are  polynomials of $p$ and $f_{\theta,i}$ and $g_{\theta,i}~(i>0)$ are polynomials of $c^2$, 
then the integrability of expansions in Eqs. \eqref{eq:seriesdifftheta} and \eqref{eq:seriesdiffphi} relies on the integrability of the following integrals 
\begin{align}
	&\int_1^{p_\mathrm{s,d}} \frac{\mathrm{polynomial}(p)}{(1+p)^{i-1}\sqrt{1-p^2}}\dd p~~(i\geq 1),\label{eq:pintgform}\\
	&\int_{c_\mathrm{s,d}}^{c_\te} \frac{\mathrm{polynomial}(c^2)}{\sqrt{c_\te^2-c^2}}\dd c.\label{eq:cintgform}
\end{align}
Fortunately, they are always integrable (see Appendix \ref{sec:appd} for the proof) and this guarantees that we can obtain a series solution for the deflection angles. 

After integration, the results for Eqs. \eqref{eq:seriesdifftheta} and \eqref{eq:seriesdiffphi} then become
\begin{align}
	& \sum_{j=s,d}\sum_{i = 1}^{\infty} F_{r,i}( p_j) \left(\frac{M}{r_0}\right)^i = \sum_{j=s,d}\sum_{i = 1}^{\infty} F_{\theta,i}(c_j,c_\te) \left(\frac{M}{r_0}\right)^i, \label{eq:Inteddifftheta}\\
	&\Delta \phi = \sum_{j=s,d} \lsb \sum_{i = 2}^{\infty} G_{r,i}( p_j)+\sum_{i = 0}^{\infty} G_{\theta,i}(c_j,c_\te) \rsb\left(\frac{M}{r_0}\right)^i .\label{eq:Inteddiffphi}
\end{align}
Here the coefficient functions $F_{r,i},\,F_{\theta,i},\,G_{r,i},\,G_{\theta,i}$ are integration results of terms containing  $f_{r,i},\,f_{\theta,i},\,g_{r,i},\,g_{\theta,i}$ respectively, and therefore are also functions of the corresponding integration limits. The first few of them, for $j=\{\mathrm{s},\,\mathrm{d}\}$, are 
\begin{subequations}
	\begin{align}\label{eq:GFdef}
		F_{r,1} =& \frac{1}{M v E}\left[\frac{\pi}{2}-\sin ^{-1}({p_j})\right],\\
		F_{r,2} =& \frac{1}{M v^3 E} \lsb\sqrt{1-{p_j}^2}\lb \frac1{1+{p_j}}+v^2\rb+\sin ^{-1}({p_j}) - \frac{\pi }{2}\rsb,\nn\\
		F_{\theta,1} =& \frac{1}{M v E}\left[\frac{\pi}{2}-\sin ^{-1}\left(\frac{c_j}{c_\te}\right)\right],\label{eq:Gtheta1}\\
		F_{\theta,2} =& \frac{1}{Mv^3E} \lsb\sin^{-1}\left(\frac{{c_j}}{c_\te}\right)- \frac{\pi }{2}\rsb,\label{eq:Gtheta2}\\
		G_{r,2}=&\frac{2 \hat{a} s_l }{v}\sqrt{1-p_i^2}-\frac{1}{2} \hat{a}^2 s_\te \left[ p_i\sqrt{1-p_i^2} +\cos^{-1}(p_i)\right],\label{eq:Gr2}\\
		G_{\theta,0} =& \frac{\pi}{4}-s_{r\theta}\tan ^{-1}  \frac{c_j s_\te}{\sqrt{c_\te^2-c_j^2}}, \label{eq:Ftheta0}\\
		G_{\theta,1} =& 0, \label{eq:Ftheta1}\\
		G_{\theta,2} =& \frac{1}{4} \hat{a}^2 s_\te \left[\frac{\pi}{2} -s_{r\theta}2 \sin ^{-1}\left(\frac{c_j}{c_\te}\right)\right].
		\label{eq:Gtheta2}
	\end{align}
\end{subequations}

One has to be careful when interpreting Eqs. \eqref{eq:Inteddifftheta} and \eqref{eq:Inteddiffphi}. Although Eq. \eqref{eq:Inteddiffphi} looks like a series of $(M/r_0)$ with $G_{r,i}$ and $G_{\theta,i}$ as coefficients for the deflection $\Delta\phi$, it is still not the true final perturbative series of $(M/r_0)$ as in the case in the equatorial plane. Reason one is that there exists dependence of $p_\mathrm{s,d}$ on $r_0$ (see Eq. \eqref{eq:auxdef}). The second and deeper reason is that, as we pointed out in the last section, among the parameters $r_{\mathrm{s,d}},\,\theta_\mathrm{s,d},\,r_0$ and $\theta_\te$, not all of them are independent. Indeed, $\theta_\mad$ can be fixed by other parameters including $r_0$, and this has to be taken into account when attempting to obtain an $(M/r_0)$ series of $\Delta\phi$. This relation can be derived from Eq. \eqref{eq:Inteddifftheta} using two methods, the perturbative method and Jacobi elliptic function method respectively. Here will directly present the result of this relation but postpone its derivation to Appendix \ref{appd:cdr0rel}
\begin{equation}
	c_\mathrm{d} \equiv \cos\theta_\mad=\sum^{\infty}_{i=0} h_i\lb \frac{M}{r_0}\rb^i, \label{eq:cdfinal}
\end{equation}
where
\begin{subequations}
	\begin{align}
		h_0 =& c_\te \cos  a_1, \label{eq:hresulth0}\\
		h_1 =& -\frac{c_\te}{v^2} a_2\sin  a_1, \label{eq:hresulth1}\\
		h_2 =&-\frac{c_\te}{4} \left[a_3\sin  a_1 + \cos  a_1\left(\frac{2}{v^4} a_2^2+\hat{a}^2  c_\te^2 \sin ^2 a_1\right)\right] \label{eq:hresulth2}
	\end{align}
\end{subequations}
and
\begin{subequations}
	\begin{align}
		a_1=&-s_{r\theta}\cos^{-1}\left(\frac{c_\mas}{c_\te}\right)+\sum_{j=s,d}\cos^{-1}\left(p_j\right),\label{eq:a1def}\\
		a_2=&\sum_{j=s,d}\left(\sqrt{\frac{1-p_j}{1+p_j}}+\sqrt{1-p_j^2}v^2\right),\\
		a_3=&s_{r\theta}\hat{a}^2 c_\mas \sqrt{c_\te^2-c_\mas^2}+\sum_{j=s,d}\left\{\left(3 -\hat{a}^2 c_\te^2\right)p_j \sqrt{1-p_j^2}\right.\nn\\
		&-\frac{8 s_l s_\te\hat{a}}{v}\frac{2+p_j}{1+p_j} \sqrt{1-p_j^2}+3(1+\frac{4}{v^2})\cos^{-1}(p_j)\nn\\
		&\left.-\frac{2}{v^2}\sqrt{\frac{1-p_j}{1+p_j}}\left[2(1+\frac{1}{v^2})+\frac{1}{v^2}\frac{p_j}{1+p_j}\right]\right\}.
	\end{align}
\end{subequations}

\subsection{The deflection angles}

To compute the deflection angle $\Delta\phi$, all we need to do is substituting Eq. \eqref{eq:cdfinal}  into \eqref{eq:Inteddiffphi}, and recollect terms involving $G_{\theta,i}$ into a power series in $(M/r_0)$ with new coefficients $G_{\theta,i}^\prime$. After this, $\Delta\phi$ finally becomes 
\begin{align}
	\Delta \phi =& \sum_{j=s,d} \lsb\sum_{i = 2}^{\infty} G_{r,i}( p_j) + \sum_{i = 0}^{\infty} G_{\theta,i}^\prime(c_\mas,c_\te)\rsb \left(\frac{M}{r_0}\right)^i \label{eq:Inteddiffphi2}
\end{align}
where $G_{r,i}$ is still given by Eq. \eqref{eq:Gr2} and
the first three orders of $G_{\theta,i}^\prime~(i=0,1,2)$ are 
\begin{align}
	G_{\theta,0}^\prime=&\pi-s_{r\theta} \left[s_{r\theta}\tan ^{-1}\left(s_\te \cot  a_1\right)+\tan ^{-1} \frac{c_\mas s_\te}{\sqrt{c_\te^2-c_\mas^2}}\right],\nn\\
	G_{\theta,1}^\prime=&\frac{ s_\te  a_2}{ \left(1-c_\te^2 \cos ^2 a_1\right)v^2},\nn\\
	G_{\theta,2}^\prime=&\frac{1}{2} \hat{a}^2 s_\te \left[ \cos^{-1}(p_\mas)+\cos^{-1}(p_\mathrm{d})\right]\nn\\
	&+\frac{s_\te }{4(1-\cos ^2(a_1) c_\te^2)} \left[-\frac{2 a_2^2 \sin (2 a_1)c_\te^2}{v^4 \left(1-\cos ^2(a_1)c_\te^2\right)}\right.\nn\\
	&\left.+a_3+\frac{1}{2}\hat{a}^2 \sin (2a_1)c_\te^2\right].
\end{align}
The null limit of this deflection can be obtained easily by taking $v=1$, resulting in, to the leading order
\begin{align}
	&\Delta\phi(v=1)=\pi- \tan ^{-1}\left(s_\te \cot  a_1\right)-s_{r\theta}\tan ^{-1}\left( \frac{c_\mas s_\te}{\sqrt{c_\te^2-c_\mas^2}}\right)\nn\\
	&+\sum_{j=s,d}\frac{ s_\te (2+p_j)\sqrt{\frac{1-p_j}{1+p_j}}}{ 1-c_\te^2 \cos ^2 a_1}\frac{M}{r_0}.
\end{align}
We have also verified that $\Delta\phi$ in Eq. \eqref{eq:Inteddiffphi2} has the correct equatorial plane limit. Specifically, if we let $\theta_\mas\to\pi/2,\,\theta_\te\to\pi/2$, this deflection angle will reduce to the result computed purely in the equatorial plane for particles with arbitrary asymptotic velocity \cite{Huang:2020trl}. 

To see the finite distance effect more clearly, we can expand $\Delta\phi$ in Eq.\eqref{eq:Inteddiffphi2} in the small $p_\mathrm{s,d}$ limit
\begin{align}
	&\Delta\phi=\sum_{n,m=0}^{n+m=2}\rho_{nm}\lb \frac{M}{r_0}\rb^n (p_\mas+p_\mathrm{d})^m+\mathcal{O}\lb \epsilon^3\rb,  \label{eq:dphismallp}
\end{align}
where $\epsilon$ stands for the infinitesimal of either $(M/r_0)$ or $p_\mathrm{s,d}$ and the coefficients are
\begin{subequations}
	\begin{align}&\rho_{00}=\pi,\\
		&\rho_{01}=-\frac{\sin(x_\mas)}{s_\mas},\\ &\rho_{02}=s_{r\theta}\frac{\sin(2x_\mas)}{2t_\mas s_\mas},\\
		&\rho_{10}=\frac{2\sin(x_\mas)}{s_\mas}\left(1+\frac{1}{v^2}\right),\\
		&\rho_{11}=-\frac{1}{s_\mas}\left[\frac{\sin(x_\mas)}{v^2}+\frac{2s_{r\theta}\sin(2x_\mas)}{t_\mas}\left(1+\frac{1}{v^2}\right)\right],\\
		&\rho_{20}=\frac{4 s_l \hat{a}  \cos(2x_\mas)}{v}+\frac{1}{s_\mas}\left\{\frac{2s_{r\theta}\sin(2x_\mas)}{t_\mas}\left(1+\frac{1}{v^2}\right)^2\right.\nn\\
		&~~~~-\sin(x_\mas)\left.\left[\left(1+\frac{1}{v^2}\right)\frac{2}{v^2}-3
		\pi  \left(\frac{1}{4}+\frac{1}{v^2}\right)\right]\right\}, \label{eq:rho20res}
	\end{align}
\end{subequations}    
where recalling $t_\mas=\tan\theta_\mas,~s_i=\sin\theta_i~(i=\mathrm{s,d})$ and we have set here and for later purpose
\begin{align}
	x_\mathrm{s,d}=\sin^{-1}\lb s_\te/s_\mathrm{s,d}\rb.
\end{align}
Note that the higher order coefficients in Eq. \eqref{eq:dphismallp} can also be found easily but are too tedious to show here. 
If we take the infinite distance limit, then this becomes 
\begin{align}
	&\Delta\phi(r_{\mathrm{s,d}}\to\infty)=\pi+\frac{2 \sin(x_\mas) \left(1+v^2\right)}{s_\mas v^2}\left(\frac{M}{r_0}\right)\nn\\
	&+\left\{\frac{4 s_l \hat{a}\cos(2x_\mas)}{v}+\frac{1}{s_\mas}\left[\frac{2s_{r\theta}\sin(2x_\mas)}{t_\mas}\left(1+\frac{1}{v^2}\right)^2\right.\right.\nn\\
	&-\sin(x_\mas)\left.\left.\left(\left(1+\frac{1}{v^2}\right)\frac{2}{v^2}-3
	\pi  \left(\frac{1}{4}+\frac{1}{v^2}\right)\right)\right]\right\}\left(\frac{M}{r_0}\right)^2\nn\\
	&+\mathcal{O}\left(\frac{M}{r_0}\right)^3.
\end{align}
If we further take the null limit of $v=1$, this simplifies to
\begin{align}
	&\Delta\phi(r_{\mathrm{s,d}}\to\infty,v= 1)=\pi+\frac{4 \sin(x_\mas)}{s_\mas}\left(\frac{M}{r_0}\right)+\left(\frac{M}{r_0}\right)^2\nn\\
	&\times\left[4 s_l \hat{a} \cos(2x_\mas) +\frac{8 s_{r\theta}\sin(2x_\mas)}{s_\mas t_\mas}+\frac{\sin(x_\mas)}{s_\mas}\left(\frac{15 \pi }{4}-4\right) \right]\nn\\
	&+\mathcal{O}\left(\frac{M}{r_0}\right)^3.
\end{align}

For the deflection in the $\theta$ direction, Eq. \eqref{eq:cdfinal} provides the desired solution for $\theta_\mad$ once $r_{\mathrm{s,d}},\,\theta_\mathrm{s,e}$ and $r_0$ are known. In other words, the deflection $\Delta\theta$ becomes
\begin{align}
	\Delta\theta=&\theta_\mad+\theta_\mas-\pi\nn\\
	=&\cos^{-1}(c_\mad)+\theta_\mas-\pi,\label{eq:dthetaf}
\end{align}
where $c_\mad$ is given by Eq. \eqref{eq:cdfinal}. 
For null rays, this deflection becomes, to the leading order
\begin{align}
	\Delta\theta(v= 1)=&\cos^{-1}\left[c_\te \cos(a_1)\right]+\theta_\mas -\pi\nn\\
	&+\sum_{j=s,d}\frac{ c_\te \left(\sqrt{\frac{1-p_j}{1+p_j}}+\sqrt{1-p_j^2}\right)}{\sqrt{ 1-c_\te^2 \cos ^2 a_1}}\frac{M}{r_0}.
\end{align}
To have a better understanding of this result, similar to the case of $\Delta\phi$, we can also expand it for small $p_\mathrm{s,d}$, and find
\begin{align}
	&\Delta\theta=\sum_{n,m=0}^{n+m=2}\tau_{nm}\lb \frac{M}{r_0}\rb^n (p_\mas+p_\mad)^m+\mathcal{O}\lb \epsilon^3\rb, \label{eq:dthetasmallp}
\end{align}
where the coefficients are
\begin{subequations}
	\begin{align}  
		&\tau_{00}=0,\\
		&\tau_{01}=-s_{r\theta}\cos(x_\mas),\label{eq:tau01res}\\
		&\tau_{02}=\frac{\cos(2x_\mas)-1}{4t_\mas},\\
		&\tau_{10}=2s_{r\theta}\left(1+\frac{1}{v^2}\right)\cos(x_\mas),\label{eq:tau10res}\\
		&\tau_{11}=\left[-s_{r\theta}\frac{\cos(x_\mas)}{v^2}+\frac{1-\cos(2x_\mas)}{t_\mas}\left(1+\frac{1}{v^2}\right)\right],\\
		&\tau_{20}=\frac{\cos(2x_\mas)-1}{t_\mas}\left(1+\frac{1}{v^2}\right)^2-\frac{4 s_{r\theta} s_l s_\mas\hat{a}\sin(2x_\mas)}{v}\nn\\
		&-s_{r\theta} \cos(x_\mas) \left[\left(1+\frac{1}{v^2}\right)\frac{2}{v^2}-3
		\pi  \left(\frac{1}{4}+\frac{1}{v^2}\right)\right]. \label{eq:tau20res}
	\end{align}
\end{subequations}
Setting $p_\mathrm{s,d}$ to zero, this yield the deflection in $\theta$ direction for source and detector at infinite radii
\begin{align} 
	&\Delta\theta(r_{\mathrm{s,d}}\to\infty)=2s_{r\theta}\left(1+\frac{1}{v^2}\right)\cos(x_\mas)\left(\frac{M}{r_0}\right)\nn\\
	&+\left\{\frac{\cos(2x_\mas)-1}{t_\mas}\left(1+\frac{1}{v^2}\right)^2-\frac{4 s_{r\theta} s_l s_\mas\hat{a} \sin(2x_\mas)}{v}\right.\nn\\
	&\left.-s_{r\theta} \cos(x_\mas) \left[\left(1+\frac{1}{v^2}\right)\frac{2}{v^2}-3
	\pi  \left(\frac{1}{4}+\frac{1}{v^2}\right)\right]\right\}\left(\frac{M}{r_0}\right)^2\nn\\
	&+O\left(\frac{M}{r_0}\right)^3.
\end{align}
Further setting $v=1$, the null limits of this deflection becomes
\begin{align} 
	&\Delta\theta(r_{\mathrm{s,d}}\to\infty,v=1)=4s_{r\theta}\cos(x_\mas)\left(\frac{M}{r_0}\right)\nn\\
	&+\left[\frac{4\left(\cos(2x_\mas)-1\right)}{t_\mas}-4s_{r\theta}  s_l \hat{a}\sin(2x_\mas)\right.\nn\\
	&\left.+s_{r\theta} \cos(x_\mas) \left(\frac{15 \pi }{4}-4\right)\right]\left(\frac{M}{r_0}\right)^2+O\left(\frac{M}{r_0}\right)^3.
\end{align}

When studying the GL due to the lens, Eqs. \eqref{eq:cdfinal} and \eqref{eq:Inteddiffphi2} allow us to solve for $(\theta_\te,\,r_0)$ once the source location $(r_\mas,\,\theta_\mas,\,\Delta\phi)$ and the detector location $(r_\mad,\,\theta_\mad,\,0)$ are fixed. Here without loss of generality, we can set the $\phi$ coordinate of the detector to zero, which means that the source $\phi$ coordinate will be $\Delta\phi$). The solutions $(r_0,\,\theta_\te)$ then can be directly used in the apparent angle formula Eq. \eqref{eq:aadefs} to yield the apparent angles of the images. 

\section{Gravitational Lensing\label{sec:gltd}}

In this section, we will demonstrate how the series solution of the deflection angles with finite distance effect taken into account can help us to solve $r_0$ and $\theta_\te$ naturally and more precisely. These quantities, in turn, lead to the desired apparent angles and their magnifications of the GL images using a set of exact formulas in Sec. \ref{subsec:aamag}.

\subsection{GL equation and solution to $r_0,\,\theta_\te$}

\begin{figure}[htp!]
	\centering
	\includegraphics[width =0.45\textwidth]{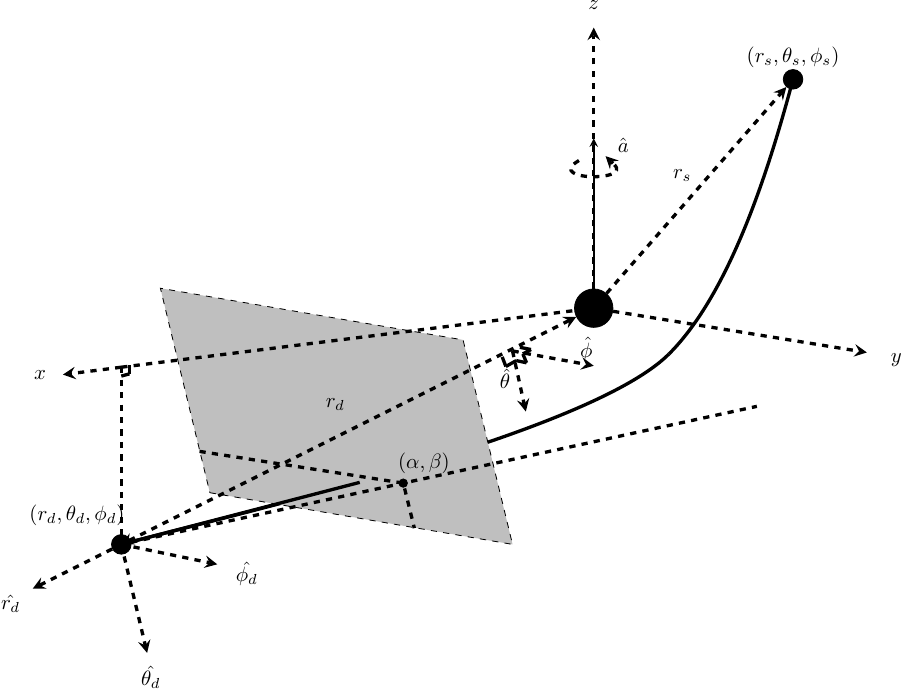}
	\caption{Schematic diagram for the deflection and lensing in the WDL in Kerr BH spacetime. The source and detectors are located at $(r_\mas,\theta_\mas,\phi_\mas)$ and $(r_\mad,\theta_\mad,\phi_\mad)$ respectively. The gray box represents a patch of the celestial plane of the detector. The apparent angles of the images on this plane are also indicated. The $(\alpha,\,\beta)$ are the small apparent angles of the image against the projected axis of $+\hat{z}$ and $+\hat{y}$. }
	\label{fig:glillustration}
\end{figure}

With the deflection in both $\theta$ and $\phi$ directions known, one can then attempt to solve for the apparent angles of the lensed images using the GL equations. Such GL equations are usually formulated using approximate geometrical relations that link the source and detector locations, the intermediate variables such as $r_0$ and $\theta_\te$, the deflection angles $\Delta\phi,\,\Delta\theta$, as well as the desired apparent angle(s), which has two components $(\alpha,\,\beta)$ in our case. In Fig. \ref{fig:glillustration}, we adopt the common setup of the detector's local inertial frame, whose basis axes consist of the detector-lens direction $\hat{r}_\mad$, and the directions $\hat{\theta}_\mad$ and $\hat{\phi}_\mad$ parallel respectively to the $\hat{\theta}$ and $\hat{\phi}$ basis of the spacetime coordinates.  

In this work, our set of the GL equation consists of two equations. The first set is nothing but the very definitions of the deflections in $\phi$ and $\theta$ directions
\begin{subequations}
	\label{eq:dphidthetaeq}
	\begin{align}
		&\Delta\phi(r_0,\theta_\te)=\phi_\mad-\phi_\mas\equiv \pi+\delta\phi,\label{eq:dphigleq}\\
		&\Delta\theta(r_0,\theta_\te)=\theta_\mad+\theta_\mas-\pi\equiv \delta\theta,
		\label{eq:dthetadef}
	\end{align}
\end{subequations}
where $\Delta\phi$ and $\Delta\theta$ are expressed as in Eqs. \eqref{eq:Inteddiffphi2} and \eqref{eq:dthetaf} and $\delta\phi$ and $\delta\theta$ are two small deviation angles characterizing the location of the source relative to the lensing-observer axis when the lens is not present. We argue that this set of GL equations is more exact because unlike many others, they are simply definitions of the deflection and their establishment requires no other geometrical approximations. Using this set of equations, once the quantities $a,\,M$ and  $(r_\mas,\,\theta_\mas,\,\phi_\mas),\,(r_\mad,\,\theta_\mad,\,\phi_\mad=2\pi)$
are given in prior, we will be able to solve for the intermediate variables, i.e., the minimal radial coordinate $r_0$ and extreme angle $\theta_\te$ that allow the test particle to reach the detector. These two quantities can also be interchanged with the other pair of kinetic variables $(L,\,E)$ of the test particle. Note that without loss of generality, we have fixed $\phi_\mad=2\pi$ and in the WDL, $\phi_\mad-\phi_\mas$ is usually very close to $\pi$. 

In practice, since Eqs. \eqref{eq:Inteddiffphi2} and \eqref{eq:dthetaf} have a more complicated dependence on $r_0$ and $\theta_\te$, when substituting into Eq. \eqref{eq:dphidthetaeq}, we will use their expanded forms, i.e. Eqs. \eqref{eq:dphismallp} and \eqref{eq:dthetasmallp} with terms of combined order higher than two truncated. Inspecting these two equations carefully, we can see that their dependence on $r_0$ can be converted to a polynomial form. 
From Eqs. \eqref{eq:rho20res} and \eqref{eq:tau20res}, we observe that the second order in the expansion of $(M/r_0)$ is also the minimal order that the effect of the spacetime spin $\hat{a}$ will appear. Therefore any attempt to study the off-equatorial plane deflection and lensing should retain the deflection angles to at least this order, which is also our approach in this work. Otherwise, the off-equatorial motion will just be a simple rotation of the equatorial motion in Schwarzschild spacetime since the spin $\hata$ is not taken into account.
For the dependence of this set of equations on $\theta_\te$ however, we see that these two equations are both linear combinations of $\sin(n x_\mas)$ and $\cos(n x_\mas)~(n=0,\,1,\,2)$, which can also be converted to polynomials of $\tan(x/2)$. Unfortunately, to the order we are interested in, these polynomials do not allow simple analytical forms for their solutions. More precisely, in order to include the effect of $\hat{a}$, we found that the $\tan(x/2)$ should be a root of an order tenth-order polynomial. Once $\tan(x/2)$ is obtained however, the $r_0$ can be simply evaluated (not solved) from a polynomial involving $\tan(x/2)$. Therefore we will use the numerical methods to solve $r_0$ and $\tan(x/2)$ in what follows.  

Besides these two equations, we also supplement a selection condition of the solutions related to the initial conditions.
When $s_{r\theta}>0$ (or equivalently $s_\theta>0$ because we always use $s_r>0$), Eq. \eqref{eq:difftheta} implies that the trajectory initially moves toward decreasing $\theta$ and therefore we will require that $\theta_e$ be smaller than $\theta_\mas$. On the contrary, if $s_{r\theta}<0$, we will require that $\theta_\te>\theta_\mas$. 

In Figs. \ref{fig:r0thetamonphis} - \ref{fig:r0thetamonhata}, we plot the solved $r_0$ and $\theta_\te$ as functions of variables $\delta\phi,\,\delta\theta,\,\theta_\mas,\,\hat{a}$ and $r_{\mathrm{s,d}}$ using Sgr A* as the central lens. We utilized its data $M=4.1\times 10^6M_\odot$ and $r_\mad=8.34$ kpc {\red \cite{EventHorizonTelescope:2022wkp}}. In principle, if we allow the spacetime spin to be negative and the locations of the source and detector to be switched, then we can restrict the non-equivalent parameter space to $\delta\phi>0,\,\delta\theta>0,\,\theta_\mas\in [0,\pi/2]$. However, to show the lensed images more comprehensively, in some of the plots in the following we will consider negative $\delta\theta,\,\delta\phi$. The choice of other parameters is provided in the caption of each plot.

Let us point out a few features of the solution process and results. First, we find that for each set of parameters and among the four different possible combinations of the signs $s_{r\theta}$ and $s_l$, there are only two combinations that allow physical solutions to $r_0$ and $\theta_\te$. 
In most of the parameter space, one of these allowed trajectories will be prograde with respect to the $+z$ axis while the other will be retrograde. In this paper, by {\it prograde} and {\it retrograde}, we specifically mean that the trajectories rotate anticlockwise and clockwise around the $+\hat{z}$ directions respectively. No retrolensing is involved because we only discuss the weak deflection cases in this work. We will denote the minimal radii and extreme $\theta$ coordinate of the prograde trajectory as $(r_{0+},\,\theta_{\te+})$ and those of the retrograde trajectory as $(r_{0-},\,\theta_{\te-})$. Since $\theta_\mas\leq \pi/2$, when $\theta_\te$ is a minimum (or maximum), the trajectory reaches the detector from above (or below) the equatorial plane after bending, as illustrated in Fig. \ref{fig:glillustration} by the two solid trajectories. 

Secondly, we would like to point out two fundamental properties of the trajectories that will help in the understanding of the results presented in many of the following figures. The first is that when $\delta\phi$ is relatively large (greater than $10^{-4\prime\prime}$) for the numerical values of other parameters we used, the spacetime spin's effect is only secondary compared to that of $\delta\phi$. This can be understood from Eqs. \eqref{eq:dphigleq} and \eqref{eq:Inteddiffphi2} that $\hat{a}$ appears one order higher than $\delta\phi$ or from the deflection angle that $\hat{a}$ appears one order higher than $M/r_0$. Under these parameter settings, then the effect of the spin can be ignored and the physics should be similar to the SSS case, in which the total deflection angle can be approximated as 
\begin{align}
	\delta\eta\approx \sqrt{(\delta\theta)^2+\sin^2\theta_\mas(\delta\phi)^2}.    \label{eq:totaldef}
\end{align} 
Then from our experience with SSS spacetime, it is known that when $\delta\phi>0$, the minimal radial coordinate $r_{0+}$ will decrease while $r_{0-}$ will increase as $\delta\eta$ increases, as in Schwarzschild spacetime \cite{Liu:2015zou}, regardless whether the increase of $\delta\eta$ is caused by the increase of $\delta\phi,\,\theta_\mas$ or $\delta\theta$.
When the effect of $\hat{a}$ is secondary, the second feature of the trajectories is that both trajectories lie essentially within a single plane that contains the source, lens and detector. With these two fundamental properties in mind, one then can study and understand more easily how various quantities affect the extreme $\theta_{\te\pm}$ by simply drawing this plane in the Cartesian coordinates. 

\subsubsection*{Effect of various parameters}

\begin{figure}[htp!]
	\centering
	\includegraphics[width=0.45\textwidth]
	{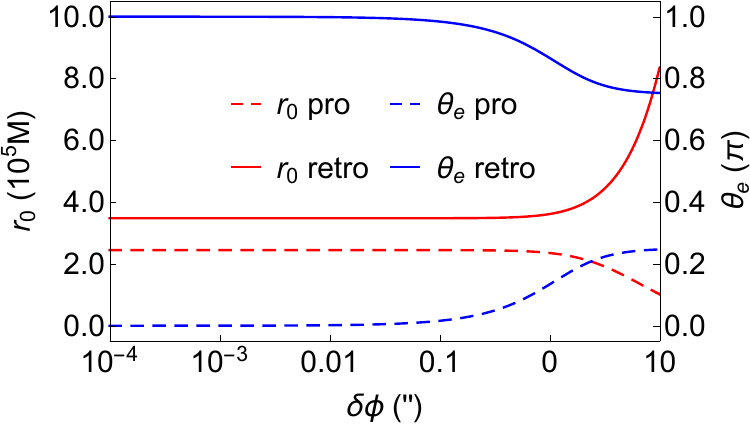}\\
	(a)\\
	\includegraphics[width=0.45\textwidth]
	{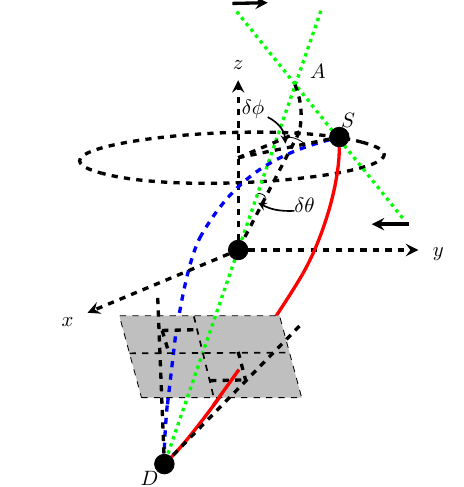}\\
	(b)\\
	\caption{(a) The dependence of $r_0$ and $\theta_\te$ on $\delta\phi$. We fixed $\theta_\mas=\pi/4,\,\delta\theta=1^{\prime\prime},\,\hat{a}=1/2,\,r_\mas=r_\mad,\,v=1$ in this plot. (b) Schematic diagram as $\delta\phi$ decreases.}
	\label{fig:r0thetamonphis}
\end{figure}

Fig. \ref{fig:r0thetamonphis} shows the dependence of $r_{0\pm}$ and $\theta_{\mathrm e\pm}$ on $\delta\phi$. This relationship is one of the main focuses of the GL in SSS or equatorial plane of stationary and axisymmetric spacetimes. As described above, for each set of fixed parameters, there are only two physical trajectories, which we denote as $(r_{0+},\,\theta_{\te+})$ for the prograde one and $(r_{0-},\,\theta_{\te-})$ for the retrograde one respectively. 
For the minimal radii, it is seen from Fig. \ref{fig:r0thetamonphis} (a) (left axis) that if the deflection $\delta\phi$ is larger than the deflection in the $\theta$ direction ($\delta\theta=1^{\prime\prime}$), then its effect to the bending of the trajectories would dominate those of the spacetime spin as well as $\delta\theta$, as can be seen from its contribution to the total deflection $\delta\eta$ in Eq. \eqref{eq:totaldef}. As $\delta\phi$ decreases, $r_{0-}$ would rapidly decrease while $r_{0+}$ will increase, which is a feature qualitatively similar to the case in the equatorial plane \cite{Huang:2020trl}.
However, as $\delta\phi$ approaches and becomes smaller than $\delta\theta$, the effect of $\delta\theta$ to the bending will fix the two minimal radii at constant values, as shown by the flat regions in the left part of Fig. \ref{fig:r0thetamonphis} (a). 
For the extreme $\theta_\te$ of the two trajectories, it is seen from Fig. \ref{fig:r0thetamonphis} (b) (right axis) that as $\delta\phi$ decreases, the  $\theta_{\te+}$ (or $\theta_{\te-}$) of the prograde (or retrograde) trajectory keeps decreasing (or increasing), indicating that the trajectory swings closer to the $z$ axis above (or below) the equatorial plane. As $\delta\phi$ becomes much smaller than $\delta\theta$, the trajectories mainly bend in the $\theta$ direction, and the $\theta_{\te+}$ and $\theta_{\te-}$ approaches $\pi$ and $0$ indefinitely. We remind the readers that the $\theta$ coordinate along the trajectory does not necessarily deviate weakly from $\theta_\mathrm{s,d}$ even in the WDL, which can be understood in the straight trajectory case in zero gravity.
Fig. \ref{fig:r0thetamonphis} (b) illustrate schematically the change of the trajectories as $\delta\phi$ decreases. 

\begin{figure}[htp!]
	\centering
	\includegraphics[width=0.45\textwidth]
	{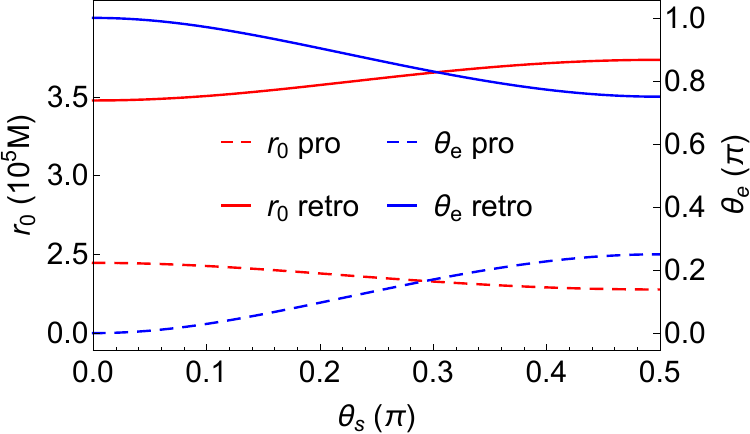}\\
	(a)\\
	\includegraphics[width=0.45\textwidth]
	{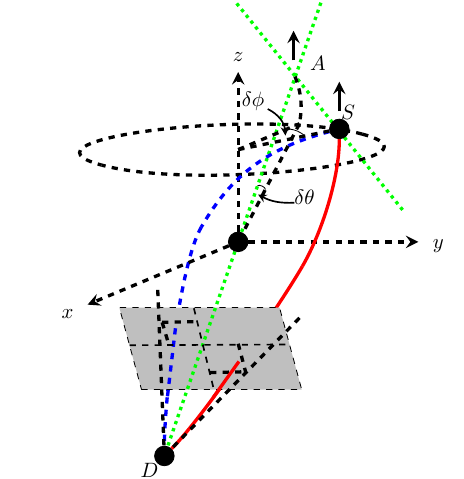}\\
	(b)\\
	\caption{(a) The dependence of $r_0$ and $\theta_\te$ on $\theta_\mas$. We fixed $\delta\phi=1^{\prime\prime},~\delta\theta=1^{\prime\prime},\,\hat{a}=1/2,\,r_\mas=r_\mad,\,v=1$ in this plot. (b) Schematic change of the trajectories as $\theta_\mathrm{s}$ decreases while keeping $\delta\theta$ and $\delta\phi$ fixed.}
	\label{fig:r0thetamonthetas}
\end{figure}

Fig. \ref{fig:r0thetamonthetas} shows how $\theta_\mas$ affects $r_{0\pm}$ and $\theta_{\te\pm}$. 
Note when adjusting $\theta_\mas$, we kept $\delta\theta=1^{\prime\prime}=\delta\phi$ a small constant so that $\theta_\mad$ is simultaneously adjusted. It is seen that first of all, compared to the effect of $\delta\phi$ on these quantities, that of $\theta_\mas$ is much weaker in general: a change of $\theta_\mas$ for about $\pi/2$ causes roughly the same amount of change for $\theta_{\te\pm}$ and a smaller change of $r_{0\pm}$ than those by a change of $\delta\phi$ for $10^{\prime\prime}$. This is however expected both from Eq. \eqref{eq:totaldef} and the fact that the approximate alignment of the source-lens-detector is not changed dramatically in the process of $\theta_\mas$'s variation.  
The second feature is that the effects of $\theta_\mas$ to both $r_{0\pm}$ and $\theta_{\te\pm}$ becomes stronger as $\theta_\mas$ decreases to zero, i.e., the $z$ axis pole directions, and weaker as it moves toward $\pi/2$, i.e., the equatorial plane. This is consistent with the first-order terms of Eq. \eqref{eq:dthetasmallp}, i.e., Eqs. \eqref{eq:tau01res} and \eqref{eq:tau10res}, which are proportional to $\cos(x_\mas)$ approaching $0$ as $\theta_\mas$ approaches the equatorial plane, and can also be seen by differentiate Eq. \eqref{eq:totaldef} with respect to $\theta_\mas$.

For the effect of $\theta_\mas$ on $r_{0\pm}$, from Fig. \ref{fig:r0thetamonthetas} (a) (left axis), we see that when the source and detectors are closer to the poles, the minimal radial coordinate $r_{0+}$ (and $r_{0-}$) for prograde (and retrograde) motion decreases (and increases) slightly. 
From our experience \cite{Liu:2015zou} with Schwarzschild spacetime with deflection $\delta\eta$ as given in Eq. \eqref{eq:totaldef}, it is not hard to anticipate that the two minimal radii should basically take the shape shown in Fig. \ref{fig:r0thetamonthetas} (a) as $\theta_\mas$ varies. 
For $\theta_{\te\pm}$, we see from Fig. \ref{fig:r0thetamonthetas} (a) (right axis) that more polar source and detector locations yield more polar $\theta_{\te\pm}$. 
This can be understood from the second property we mentioned above that each trajectory lies basically in one plane containing the source, lens and detector. One can show by drawing this plane in the Cartesian coordinates that the closer the $\theta_\mas$ to the $+z$-axis, the closer the $\theta_{\te\pm}$ to the poles.
The change caused by the variation of $\theta_\mas$ is schematically shown in Fig.\ref{fig:r0thetamonthetas} (b). 

\begin{figure}[htp!]
	\centering
	\includegraphics[width=0.45\textwidth]
	{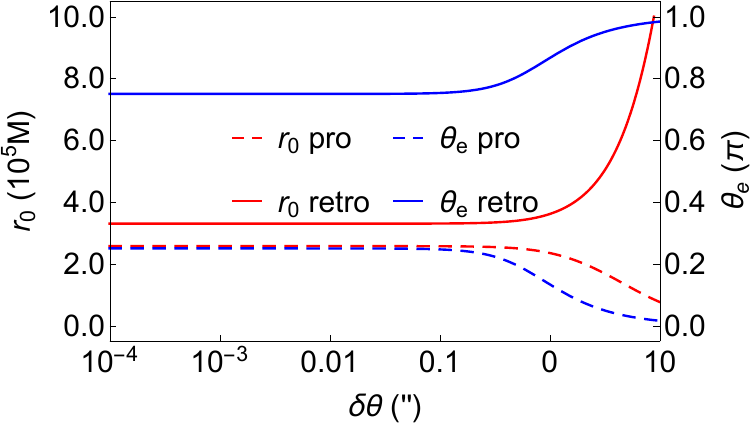}\\
	(a)\\
	\includegraphics[width=0.45\textwidth]
	{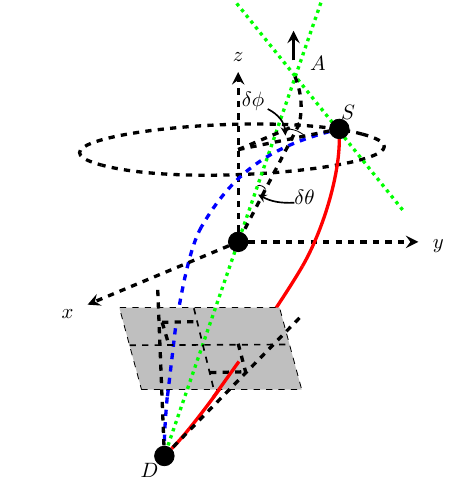}\\
	(b)\\
	\caption{(a) The dependence of $r_0$ and $\theta_\te$ on $\delta\theta$. We fixed $\delta\phi=1^{\prime\prime},\,\theta_\mas=\pi/4,\,\hat{a}=1/2,\,r_\mas=r_\mad,\,v=1$ in this plot.  (b) Schematic diagram as $\delta\theta$ varies.  }
	\label{fig:r0thetamondeltatheta}
\end{figure}

The effect of $\delta\theta$ on $r_{0\pm}$ and $\theta_{\te\pm}$, as shown in Fig. \ref{fig:r0thetamondeltatheta}, is related to the effects of $\theta_\mas$ itself in Fig. \ref{fig:r0thetamonthetas} and $\delta\phi$ in Fig. \ref{fig:r0thetamonphis} through the combination of these three parameters into the total deflection, as in Eq. \eqref{eq:totaldef}. From Fig. \ref{fig:r0thetamondeltatheta} (a) (left axis) we see that, as $\delta\theta$ increases to about $10^{\prime\prime}$, the $r_{0+}$ for prograde trajectory increases while $r_{0-}$ for retrograde trajectory decreases. The amount of their changes however is larger than those in Fig. \ref{fig:r0thetamonphis} (a) because of the extra factor of $\sin^2\theta_\mas=1/2$ in Eq. \eqref{eq:totaldef}.
The  difference between the effect of $\delta\theta$ and $\delta\phi$ appears in their effects to   $\theta_{\te\pm}$. The increase of $\delta\theta$ with a fixed $\theta_\mas$ means the increase of $\theta_\mad$. Therefore for an increasing $\delta\theta$ but a fixed $\delta\phi$, one can find that the plane containing the source, lens, detector and the two trajectories will be tilted more vertically towards the $z$-axis. When $\delta\phi>0$, this effectively increases $\theta_{\te-}$ and decreases $\theta_{\te+}$, as seen from Fig. \ref{fig:r0thetamondeltatheta} (a) (right axis). The variation of the trajectories with the increase of $\delta\theta$ is schematically shown in Fig. \ref{fig:r0thetamondeltatheta} (b). 

\begin{figure}[htp!]
	\centering
	\includegraphics[width=0.4\textwidth]
	{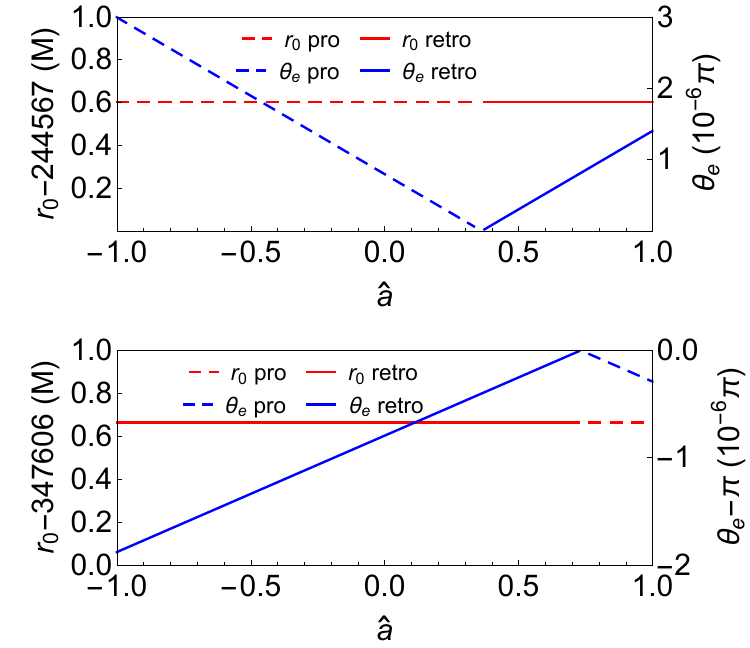}\\
	(a)\\   
	\includegraphics[width=0.4\textwidth]
	{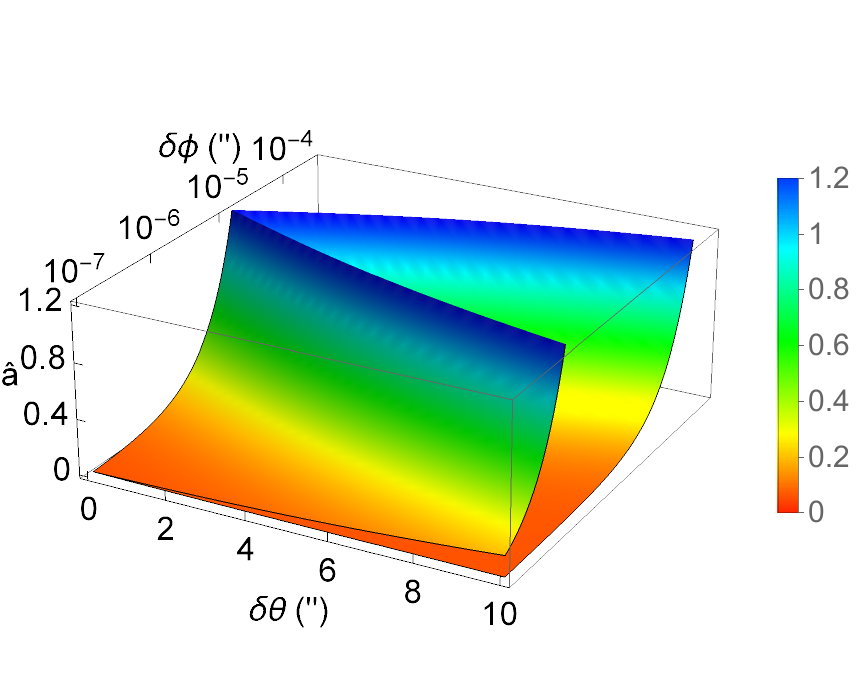}\\
	(b)
	\caption{(a) The dependence of $r_0$ and $\theta_{\te\pm}$  on $\hat{a}$. We fix $\delta\theta=1^{\prime\prime}, \delta\phi=5*10^{-6\prime\prime}, \theta_\mas=\pi/4,\,r_\mas=r_\mad,\,v=1$. (b) The critical $\hat{a}_{c\pm}$. We fixed $\theta_\mas=\pi/4,\,r_\mas=r_\mad,\,v=1$ in this plot. }
	\label{fig:r0thetamonhata}
\end{figure}

Finally we plotted how the spacetime spin $\hat{a}$ affects the $r_{0\pm}$ and $\theta_{\te\pm}$ in Fig. \ref{fig:r0thetamonhata}. 
As mentioned before, we found that when the total deflection $\delta\eta$ is larger than a certain value (roughly $\sim 10^{-5\prime\prime}$), the effect of $\hat{a}$ on these quantities are so weak that no noticeable changes are seen in these plots.
Therefore we decreased in these figures $\delta\phi$ and $\delta\theta$ simultaneously from $10^{-5\prime\prime}$ to about $10^{-7\prime\prime}$. As the deflections decrease, the influence of $\hat{a}$ starts to appear.

One of the most remarkable features that we observe in Fig. \ref{fig:r0thetamonhata} (a), is that when $\delta\phi$ is small, there exists a transition between trajectories with different choices of $(s_{r\theta},s_l)$ when $\hat{a}$ passes some critical values. When the $\delta\phi$ become sub-$\sim 10^{-5\prime\prime}$, the 
solution with $(s_{r\theta}=+1,s_l=+1)$ 
cease to exist when $\hat{a}$ is larger than $\hat{a}_{c+}$
and solution with  $(s_{r\theta}=-1,s_l=-1)$ cease to exist when $\hat{a}$ is larger than another critical value $\hat{a}_{c-}$. Instead, the above two solutions switch their sign choices to $(s_{r\theta}=+1,s_l=-1)$ and $(s_{r\theta}=-1,s_l=+1)$ respectively. This means for $\delta\phi>0,\,\delta\theta>0$ and $\theta_\mas<\pi/2$, the test particle reaching the detector from bending above (or below) the equatorial plane switches from prograde to retrograde (or from retrograde to prograde). In other words, the two trajectories intersect with the $z$ axis at $\hat{a}=\hat{a}_{c+}$ and $\hat{a}=\hat{a}_{c-}$ respectively and therefore $\theta_{\te\pm}=0,\,\pi$ as confirmed in Fig. \ref{fig:r0thetamonhata} (a).

This observation effectively provides us with the following criterion to solve for $\hat{a}_{c\pm}$: $\theta_e=0~\text{or}~\pi$. Substituting this into the lensing Eqs. \eqref{eq:dphidthetaeq}, we are able to solve for the critical $\hat{a}_{c\pm}$ to the leading order, as a function of $\delta\theta,\,\delta\phi$ and spacetime parameters.
In Fig. \ref{fig:r0thetamonhata} (b), we plot the exact dependence of $\hat{a}_{c\pm}$ on $\delta\theta$ and $\delta\phi$ while holding other parameters such as $\theta_\mas,\,r_{\mathrm{s,d}}$ etc. 
It is seen that the smaller the $\delta\phi$, the smaller the transition spins $|\hat{a}_{c\pm}|$, indicating this switching of the signs is mainly a spacetime spin effect. 
Moreover, if the spacetime is a BH one ($|\hat{a}|\leq 1$), then only for small $\delta\phi$, there exists critical $\hat{a}_{c\pm}$. For $\hat{a}$ below (or above) these two surfaces, the trajectories as shown in Fig. \ref{fig:r0thetamonthetas} - Fig. \ref{fig:r0thetamondeltatheta} with signs $(s_{r\theta}=+1,s_l=+1)$ and $(s_{r\theta}=-1,s_l=-1)$ (or $(s_{r\theta}=+1,s_l=-1)$ and $(s_{r\theta}=-1,s_l=+1)$) are the physical solutions respectively. 

There are a few other features worth remarking upon these transitions. Firstly, we see from this plot that the transitions depend on $\delta\theta$ much more weakly than on $\delta\phi$. This is understandable because $\hat{a}$ is along the direction around which $\phi$ coordinate evolves, but not along the direction of the $\theta$ coordinate. This is also consistent with our knowledge about the deflections in the equatorial plane, where the effect of spin $\hat{a}$ is most apparent only when $\delta\phi$ is very small \cite{Liu:2020mkf}. Secondly, we also note that when $\delta\phi$ is small and fixed, the spin effect is stronger for larger $\delta\theta$ in that the corresponding $\hat{a}_c$ is smaller. 
Thirdly, these transitions can also be viewed as caused by the variation of $\delta\theta$ or $\delta\phi$ when other parameters are fixed. That is, if we fix a constant $\hat{a}$, then for each $\delta\theta$ there exist critical $\delta\phi$'s below which the sign choice for $(s_{r\theta},\,s_l)$ would be $(+,-)$ and $(-,+)$. 
Lastly, it is seen that there exist some ranges of $\delta\theta$ and $\delta\phi$ in which the $\hat{a}_{c\pm}$ can exceed the extreme Kerr BH limit of 1, and therefore for these deflection angles the transition will not happen if we only consider the BH spacetime case. However, even for the Kerr spacetime with a naked singularity (the $\hat{a}>1$ part in this plot), we emphasize that the critical $\hat{a}_{c\pm}$ still exists and our plot is still valid. 

For $r_{0\pm}$, previously, it was known that in the equatorial plane, an increase in $\hat{a}$ will decrease the $r_{0+}$ and increase the $r_{0-}$ \cite{Liu:2020mkf}. We notice from the zoomed-in figure that this trend is qualitatively unchanged in the off-equatorial plane case, and it will be more apparent for very small $\delta\theta$ and $\delta\phi$ (e.g. $\sim 10^{-7\prime\prime}$). 
The influence of $\hat{a}$ on $\theta_{\te\pm}$ although seems weak in size, is actually more interesting than its effect on $r_{0\pm}$. 

\subsection{The apparent angles and magnifications \label{subsec:aamag}}

\subsubsection*{Apparent angles}

Once $(r_0,\,\theta_\te)$ or $(L,E)$ are solved for a given set of small $\delta\theta$ and $\delta\phi$, previously based on some approximate geometrical relations, Refs. \cite{Bray:1985ew,Kraniotis:2010gx} have developed approximate formulas for the apparent angles of the images observed by a static observer at $(r_\mad,\,\theta_\mad,\,\phi_\mad)$.
However, here we will use the following exact definition of the apparent angles derived from the projection of the test particle trajectory onto the celestrial sphere of the observer (see Fig. 
\ref{fig:glillustration} for the meaning of these small angles)
\begin{subequations}
\label{eq:aadefs}
\begin{align}
	&\alpha=\sin^{-1} \frac{L(\Delta_\mad-a^2 s_\mad^2)+2 a M \E  r_\mad s_\mad^2}{s_\mad\sqrt{\Delta_\mad\Sigma_\mad(\E ^2\Sigma_\mad-m^2(\Delta_\mad-a^2 s_\mad^2))}},\label{eq:aaalphares}\\
	&\beta=\sin^{-1} \frac{s_{r\theta}\sqrt{\Theta(c_\mad)(\Delta_\mad-a^2 s_\mad^2)}}{s_\mad\sqrt{\Sigma_\mad\lsb \E ^2\Sigma_\mad-m^2(\Delta_\mad-a^2 s_\mad^2)\rsb}}.\label{eq:aabetares}
\end{align}
\end{subequations}
Substituting $(\Theta(c_\mad),\,L,\,E)$ using Eqs. \eqref{eq:Thetadef}, \eqref{eq:ltor0}, \eqref{eq:ktor0} and further expanding as series of $M/r_0$ and $r_0/r_\mad)$, they can be transformed to
\begin{subequations}
\label{eq:alphabetaeq}
\begin{align}
	\alpha_\pm=&s_l  \sin (x_{\mad\pm})\lcb \frac{r_0}{r_\mad}+\frac{M }{r_\mad v^2}-\frac{M r_0}{r_\mad^2 v^2 }\right.\nn\\
	&+\frac{1}{2 {r_0} {r_\mad}} \left[a^2 -\frac{4s_ls_{\te\pm} a M}{v}+\frac{M^2}{v^4} \left(4 v^2-1\right)\right]\nn\\
	&\left.+\frac{r_0^3  \sin ^2(x_{\mad\pm})}{6 {r_\mad}^3 }\rcb,\label{eq:aaalpharesx}\\
	\beta_\pm=&
 s_{r\theta}\cos (x_{\mad\pm})\lcb \frac{r_0}{r_\mad}+\frac{ M}{r_\mad v^2}-\frac{Mr_0 }{r_\mad^2 v^2}\right.\nn\\
	&+\frac{1}{2
		r_0 r_\mad } \left[c_{\mad\pm}^2a^2  -\frac{4 s_ls_{\te\pm}a M }{v} +\frac{M^2}{v^4} \left(4 v^2-1\right)\right]\nn\\
	&\left.+\frac{r_0^3 \cos ^2(x_{\mad\pm})}{6 {r_\mad}^3 }\rcb,\label{eq:aabetaresx}
\end{align} 
\end{subequations}
where we recall that 
\begin{align} 
x_{\mad\pm}=\sin^{-1}\lb \frac{s_{\mathrm{e}\pm}}{s_{\mad\pm}}\rb=\sin^{-1}\lb \frac{\sin(\theta_{\mathrm{e}\pm})}{\sin(\theta_\mad(r_{0\pm},\theta_{\mathrm{e}\pm}))}\rb.
\end{align}
For the derivation of these formulas, see Appendix \ref{sec:appdappang}. Note that the $r_{0\pm}$ and $\theta_{\te\pm}$ enter these apparent angles through the angular momentum $L$ and the Carter constant $K$, which appears in $\Theta(c_\mad)$. These formulas have the advantage that they are applicable regardless of whether the test particle was bent weakly or strongly, although we only focus on the former case in this work. 
We have also verified in the weak deflection and equatorial plane limit that $\beta$ approaches $0$ and $\alpha$ yields the corresponding results in Ref. \cite{Huang:2020trl} (after switching from impact parameter to $r_0$). 
Moreover, if we are interested in a single apparent angle $\gamma$ between the test particle and the direction of the Kerr BH, this is given by Eq. \eqref{eq:anglegammaphi}
\begin{align}
	\gamma_\pm=&\cos^{-1} \left\{ \frac{[(a L_\pm-(a^2+r_\mad^2)\E )^2-(K_\pm+m^2 r_\mad^2)\Delta_\mad]}{\Delta_{\mad\pm}\Sigma_{\mad\pm}\lsb \E ^2\Sigma_{\mad\pm}-m^2(\Delta_{\mad\pm}-a^2 s_{\mad\pm}^2)\rsb}\right.\nn\\
	&~~~~~~~~\times (\Delta_{\mad\pm}-a^2s_{\mad\pm}^2)\bigg\}^{1/2}.\label{eq:anglegammarfront}
\end{align}

\begin{figure}[htp!]
	\centering
	\includegraphics[width=0.45\textwidth]{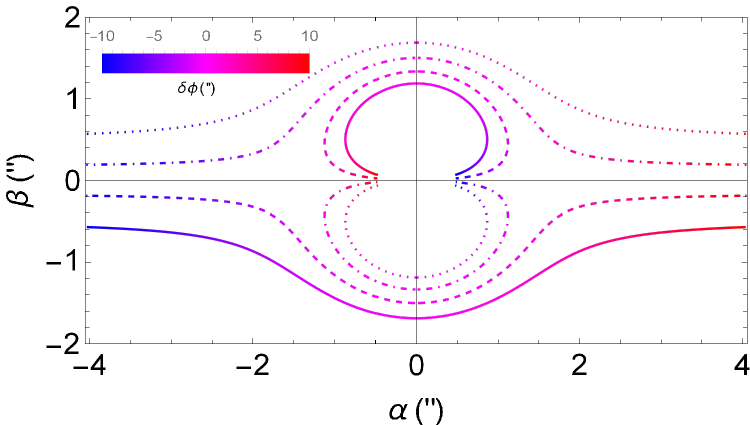}\\
	(a)\\
	\includegraphics[width=0.2\textwidth]{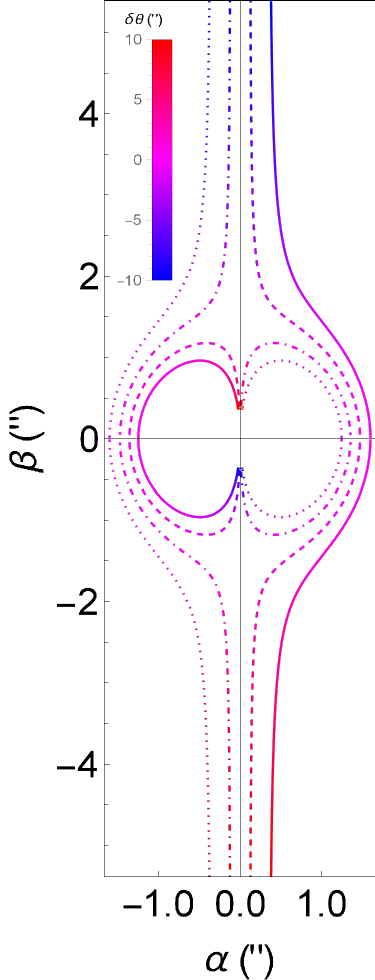}\hspace*{5cm} \\
	\hspace*{-4cm}(b)\\
	\vspace*{-10.5cm} \hspace*{3.5cm} \includegraphics[width=0.25\textwidth]{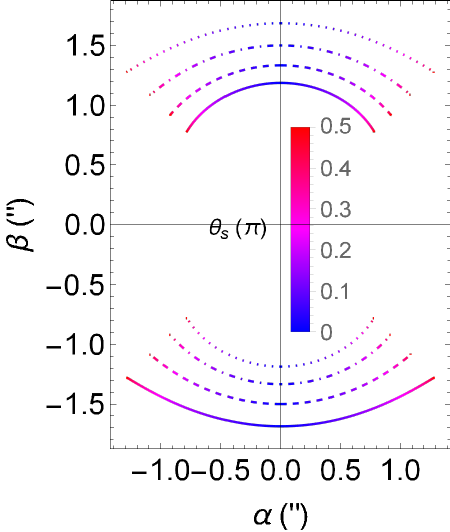}\\
	\hspace*{4cm}(c)\\
	\hspace*{3.5cm}
	\includegraphics[width=0.25\textwidth]{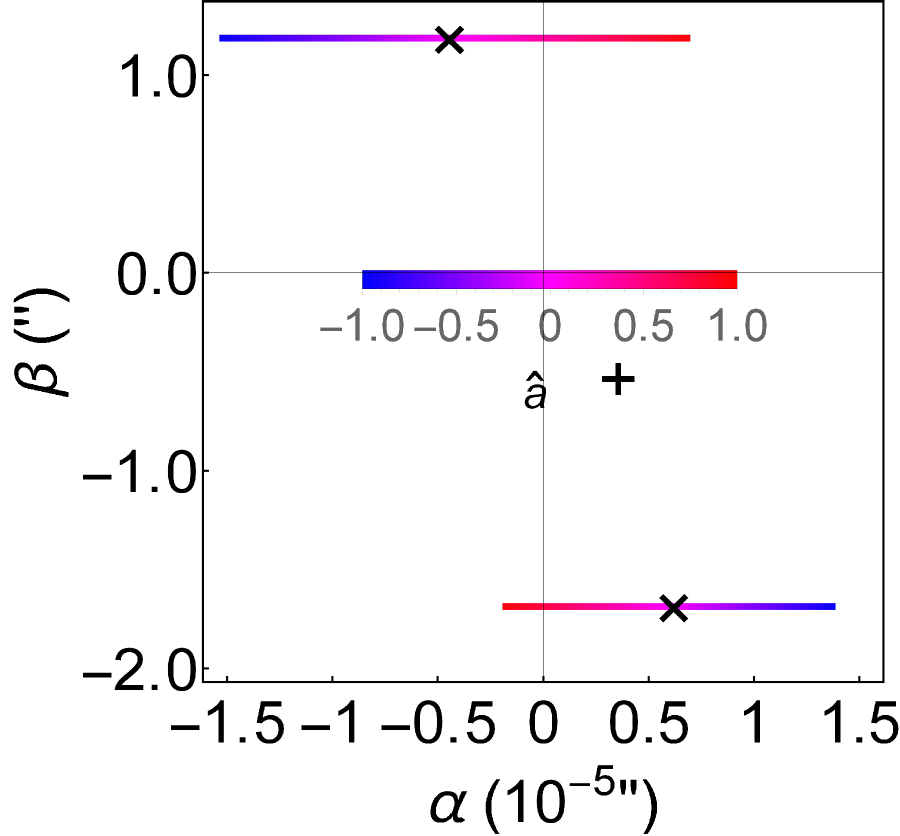}\\
	\hspace*{4cm}(d)
	\caption{The dependence of the apparent angles $(\alpha,\,\beta)$ on $\delta\phi$ from $-10^{\prime\prime}$ to $10^{\prime\prime}$ (a), on $\delta\theta$ from $-10^{\prime\prime}$ to $10^{\prime\prime}$ (b), and on $\theta_\mas$ from $0.01\pi$ to $\pi/2$ (c), and $\hat{a}$ from -1 to 1 in (d) for $\delta\phi=5\times 10^{-6\prime\prime}$ and $\delta\theta=1^{\prime\prime}$. Different line types in (a), (b), (c) are for different values of $\delta\theta,\,\delta\phi$ and $\delta\theta=\delta\phi$. The solid, dashed, dot-dashed and dotted lines represent  $1^{\prime\prime},\,0.33^{\prime\prime},\,-0.33^{\prime\prime},\,-1^{\prime\prime}$ respectively. The default values of parameters in each subplot, except those varied, are $\delta\theta=1^{\prime\prime},\,\delta\phi=1^{\prime\prime},\,\hat{a}=1/2,\,r_\mas=r_\mad,\,v=1$. }
	\label{fig:aaonparas}
\end{figure}

To reveal the dependence of the image positions on the parameters $\delta\phi,\,\theta_\mas,\,\delta\theta$ and $\hat{a}$, in Figs. \ref{fig:aaonparas}, we plot the angular locations $(\alpha_\pm,\,\beta_\pm)$ using Eq. \eqref{eq:alphabetaeq} of the prograde and retrograde images formed by trajectories with $(r_{0\pm},\,\theta_{\te\pm})$ in the celestial plane shown in Fig. \ref{fig:glillustration}. Note that if the lens were absent, it would be straightforward to determine that the source would appear to be at the point \begin{align} (\alpha,~\beta)=\lb\frac{r_\mas\sin\theta_\mas\delta\phi}{r_\mas+r_\mad},~-\frac{r_\mas\delta\theta}{r_\mas+r_\mad}\rb \label{eq:ababsent}
\end{align} on the celestial plane of the observer.

In Fig. \ref{fig:aaonparas} (a), we vary continuously the $\phi$ coordinate and consequently $\delta\phi$ of the source while keeping $\theta_\mas$ and $\hat{a}$ fixed and show the tracks of the images for several discrete $\delta\theta$. The value of $\delta\phi$ is color-coded so that basically the left side of the tracks corresponding to $\delta\theta=-10^{\prime\prime}$ and the right side to $\delta\theta=10^{\prime\prime}$.
For each fixed set of parameters in the chosen parameter range, there always come two conjugate images distributed in opposite quadrants, on the same straight line passing the origin. For $\hat{a}=1/2>0$ and $\delta\theta>0$, the image pairs in the first and third (or the fourth and second) quadrants are when $\delta\phi<0$ (or $\delta\phi>0$) and correspond to retrograde and prograde test particles respectively. For $\hat{a}=1/2>0$ and $\delta\theta<0$, the opposite happens. 
In each pair of images, the one on the outer curves, i.e. the curves further away from the origin, has a larger minimal radial coordinate and the one on the inner circular curves has a smaller $r_0$. Since for the chosen parameter ranges of $\delta\theta$ and $\delta\phi$, the effect of $\hat{a}$ is not apparent (see Fig. \ref{fig:r0thetamonhata}), the lens images look almost symmetric for $\delta\theta$ or $\delta\phi$ with opposite signs. 

Among each pair of the images,
it is seen that for each fixed $\delta\theta$, with the decreasing of $\delta\phi$, the $\alpha$-coordinate of the far-side image increases monotonically. Again, the reason is simply that the variation of the $\phi$ coordinate of the source is indeed parallel to the $\alpha$-axis in the celestial plane. The qualitative features of the angular locations of the inner images are actually more interesting. When $\delta\phi$ increases to large values, 
although the $\beta$ coordinates of the outer images do not approach zero, those of the inner images do. While when $|\delta\phi|$ decreases from large values, the $\alpha$ coordinates of the inner images do not decrease monotonically, but increase to a maximal value first and then decrease, indicating that the effect of $\delta\phi$ starts to dominate the image locations. This last feature is actually in accord with Fig. \ref{fig:r0thetamonphis} (c). 

Fig. \ref{fig:aaonparas} (b) shows the dependence of the image locations on $\delta\theta$ for a few fixed $\delta\phi$.
Qualitatively, this figure resembles a rotation of Fig. \ref{fig:aaonparas} (a), indicating that the role of $\delta\phi$ in Fig. \ref{fig:aaonparas} (a) is now played by $\delta\theta$. An apparent difference is that the range of the $\beta$ angle in Fig. \ref{fig:aaonparas} (b) is about $1.4\approx \sqrt{2}=1/\sin\theta_\mas$ times that of the $\alpha$ angle in Fig. \ref{fig:aaonparas} (a). This is a reflection of the $\sin\theta_\mas$ factor in the contribution of $\delta\theta$ and $\delta\phi$ to the total deflection in Eq. \eqref{eq:totaldef}.

In Fig. \ref{fig:aaonparas} (c) we illustrate the effect of $\theta_\mas$ on the image location while fixing $\delta\phi=\delta\theta=1^{\prime\prime}$.
It is seen that when $\theta_\mas$ approaches the spacetime rotation axis, both images are shifted to very close to the $\alpha$ axis too, which is consistent with the fact that both $\theta_{\te\pm}$ approach the $\hat{z}$ axis in Fig. \ref{fig:r0thetamonthetas} (b). When $\theta_\mas$ turns to $\pi/2$ however, the images do not approach zero $\beta$ but rather some finite $\beta$ that is smaller than $\delta\theta$. This is also in accord with the observation from Fig. \ref{fig:r0thetamonthetas} (b) that the $\theta_{\te\pm}$ approach only some middle values not close to either 0 or $\pi/2$. The more fundamental reason for these phenomena is simply that $\delta\theta$ is still non-zero in this case, i.e., the detector is still below the equatorial plane.  In general, the apparent angle $\gamma_\pm\approx \sqrt{\alpha_\pm^2+\beta_\pm^2}$ in this figure is not changed much as $\theta_\mas$ varies, because in this case the total effective deflection given by Eq. \eqref{eq:totaldef} is not changed by a large factor.

For the ranges of parameters considered in Fig. \ref{fig:aaonparas} (a) - (c), it is actually easy to verify that the critical situation in which the effect of $\hat{a}$
becomes significant is never reached. Therefore for these parameter ranges, in principle, we expect that the apparent angles can be well approximated by the results in Schwarzschild spacetime. There, when the source is located on the equatorial plane, the apparent angles against the lens-detector axis to the leading order are \cite{Xu:2021rld}
\begin{align}
\alpha_{S,\pm}=& \frac{r_\mas \sqrt{\delta\theta^2+\sin^2\theta_\mas\delta\phi^2}}{2(r_\mas+r_\mad)}\lb\text{sgn}(\delta\phi)\mp \zeta\rb,\label{eq:schimg}\end{align}
where 
\begin{align}
\zeta=&\sqrt{1+\frac{8M(r_\mad+r_\mas)\lb 1+\frac{1}{v^2}\rb}{ r_\mad r_\mas(\delta\theta^2+\sin^2\theta_\mas\delta\phi^2)}}.\label{eq:schimg2}
\end{align}
Here we have replaced the $\delta\phi$ in the Schwarzschild spacetime with the total deflection $\delta\eta$ in the Kerr spacetime, i.e., Eq. \eqref{eq:totaldef}.
However, since the source now is not located on the equatorial plane, these images should be rotated on the celestial plane such that the trajectories are in the same plane as the source, lens and detector. The location of the source with the absence of the lens in Eq. \eqref{eq:ababsent} provides for the two images a rotation angle $\xi$ from the $+\hat{\alpha}$ axis on the celestial plane, i.e.,
\begin{align}
	\cos\xi=\sin\theta_\mas\delta\phi/\delta\eta,~\text{and}~\sin\xi=-\delta\theta/\delta\eta.
\end{align}
Applying the above rotation to the apparent angles in Eq. \eqref{eq:schimg}, we finally find the apparent angles $(\alpha_\pm,\,\beta_\pm)$ for sources in Kerr spacetimes with a small $\hata$ as
\begin{align}
	\alpha_\pm=&\frac{\sin\theta_\mas \delta\phi\, r_\mas}{2(r_\mas+r_\mad)}\lb 1\mp \text{sgn}(\delta\phi)\zeta\rb, \\
	\beta_\pm=& \frac{- \delta\theta\, r_\mas}{2(r_\mas+r_\mad)}\lb 1\mp \text{sgn}(\delta\phi)\zeta\rb,
\end{align}
where $\zeta$ is in Eq. \eqref{eq:schimg2}.
We redrew the image locations $(\alpha_\pm,\,\beta_\pm)$ using the equations provided above for parameters given in the caption of Fig. \ref{fig:aaonparas} (a) - (c) and found excellent agreement with these figures. Besides, these formulas also can be used to explain the relevant results in Ref. \cite{Sereno:2003kx}. 

Fig. \ref{fig:aaonparas} (d) shows the effect of $\hat{a}$ on the apparent angles of the images. We intentionally choose small but positive $\delta\phi$ so that $\hat{a}$ can pass the critical $\hat{a}_c$ discussed in Fig. \ref{fig:r0thetamonhata} as it varies from $-1$ to 1. 
The two black crosses mark the images for $\hat{a}=0$ and the plus sign marks the location of the source if the lens were absent.
What is most interesting in these plots, and in contrast to the cases in (a) - (c) where the effect of $\hat{a}$ is not evident or equivalently the Schwarzschild case, is the feature that as $\hat{a}$ increase from zero, the retrograde (or prograde) trajectories starts to approach the $+\hat{z}$ axis (or the $-\hat{z}$ axis) and the corresponding image starts to approach the $\hat{\beta}$ axis from left (or from right). When $\hat{a}$ passes $\hat{a}_{c+}$, the retrograde trajectory intersects the $\hat{z}$ axis first and then its image appears on the right side of the $\hat{\beta}$ axis. In other words, until $\hat{a}$ reaches $\hat{a}_{c-}$, there are two prograde trajectories and images with $\alpha>0$. Eventually, when $\hat{a}$ passes $\hat{a}_{c-}$, the initially retrograde trajectory passes the $-\hat{z}$ axis and yields the image on the left side of the $\hat{\beta}$ axis. There will be one image from the prograde trajectory and one image from the retrograde trajectory again. 

\subsubsection*{Magnifications}

The magnification of the images is defined as the ratio between the observed image angular size to the source angular size if the lens was absent
\begin{align}
	\mu_\pm=&\frac{\dd \Omega_i }{\dd \Omega_i^\prime}= \frac{(r_\mad+r_\mas)^2}{r_\mas^2\sin\theta_{\mad\pm}} J_\pm\nn\\
	=&\frac{(r_\mad+r_\mas)^2}{r_\mas^2\sin\theta_\mad} 
	\left|\displaystyle\begin{array}{cc} \frac{\partial \alpha_\pm}{\partial (\delta\theta)} &
		\frac{\partial \alpha_\pm}{\partial (\delta\phi)}\\
		\frac{\partial \beta_\pm}{\partial (\delta\theta)} &
		\frac{\partial \beta_\pm}{\partial (\delta\phi)}\end{array}
	\right|,
\end{align}
where $J_\pm$ is the Jacobian of the transformation from variables $(\delta\theta,\,\delta\phi)$ to $(\alpha_\pm,\,\beta_\pm)$. 
This agrees with Ref. \cite{Wei:2011nj} which considered the (quasi-)equatorial plane case. 

Using the Eqs. \eqref{eq:aaalphares} and \eqref{eq:aabetares}, the $(\alpha_\pm,\,\beta_\pm)$ are related to $(L,\Theta)$, which according to Eqs. \eqref{eq:Thetadef} and then \eqref{eq:ltor0} and \eqref{eq:ktor0}, are functions $(r_{0\pm},\,\theta_{\te\pm})$. These quantities can be finally connected to $\delta\theta$ and $\delta\phi$ through solutions of $\theta_{\te\pm}$ and $r_{0\pm}$.
Therefore, using the chain rule for the partial derivatives, each element in the Jacobian can be computed as
\begin{subequations}
	\begin{align}
\frac{\partial\alpha}{\partial(\delta y)}=& 
\frac{\dd \alpha}{\dd L}\lb \frac{\partial L}{\partial r_0} \frac{\partial r_0}{\partial(\delta y)} + \frac{\partial L}{\partial \theta_\te} \frac{\partial \theta_\te}{\partial(\delta y)}\rb,~y\in\{\theta,\phi\},\\
\frac{\partial\beta}{\partial(\delta y)}=&
\frac{\dd \beta}{\dd \Theta}\lsb \lb\frac{\partial \Theta}{\partial L}+\frac{\partial \Theta}{\partial K}\frac{\partial K}{\partial L}\rb \lb \frac{\partial L}{\partial r_0} \frac{\partial r_0}{\partial(\delta y)} + \frac{\partial L}{\partial \theta_\te} \frac{\partial \theta_\te}{\partial(\delta y)}\rb
  \right.\nn\\
&\left.+\frac{\partial \Theta}{\partial K} \frac{\partial K}{\partial \theta_\te} \frac{\partial \theta_\te}{\partial(\delta y)} \rsb,
	\end{align}
\end{subequations}
where $y$ can be either $\theta$ or $\phi$.
Substituting $r_{0\pm}$ and $\theta_{\te\pm}$ for each image into the above equation, one can immediately obtain the magnifications of the two images. We will denote the magnification for the prograde image as $\mu_+$ and for the retrograde image as $\mu_-$. 

When $\delta\theta$ is large so that the effect of $\hat{a}$ is weak, then this magnification can be simplified to that of a Schwarzschild spacetime 
\begin{align}
	\mu_\pm=&\frac{u^2+2}{2u\sqrt{u^2+4}}\mp\text{sgn}(\delta\phi) \frac{1}{2},\label{eq:schmag}\\
	\text{where}~u=& \sqrt{\frac{r_\mad(r_\mas+r_\mad)(\delta\theta^2+\sin^2\theta_\mas\delta\phi^2)}{2Mr_\mas\lb 1+\frac{1}{v^2}\rb}}
\end{align}
and the total deflection $\delta\eta$ in Eq. \eqref{eq:totaldef} has replaced the corresponding deflection $\delta\phi$ in the  Schwarzschild spacetime. 
From this, the effect of parameters 
$\delta\phi,\,\delta\theta,\,\theta_\mas$ are very apparent.

\begin{figure}[htp!]
	\centering
	\includegraphics[width=0.45\textwidth]
	{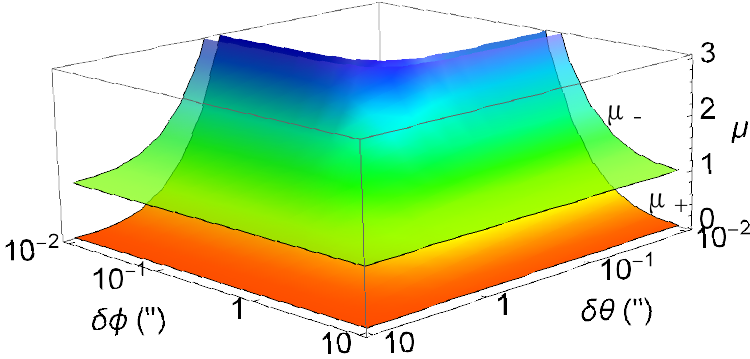}\\
	(a)\\
	\includegraphics[width=0.45\textwidth]
	{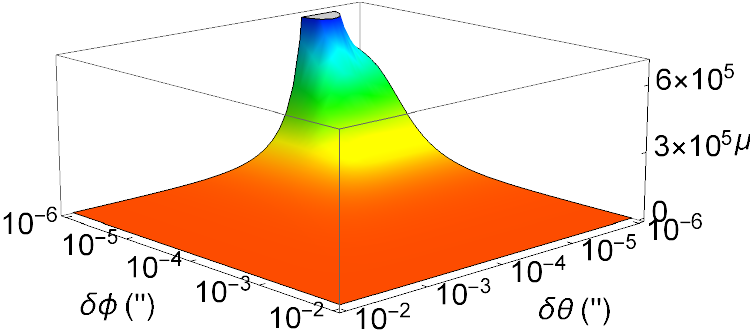}\\
	(b)\\
	\includegraphics[width=0.45\textwidth]
	{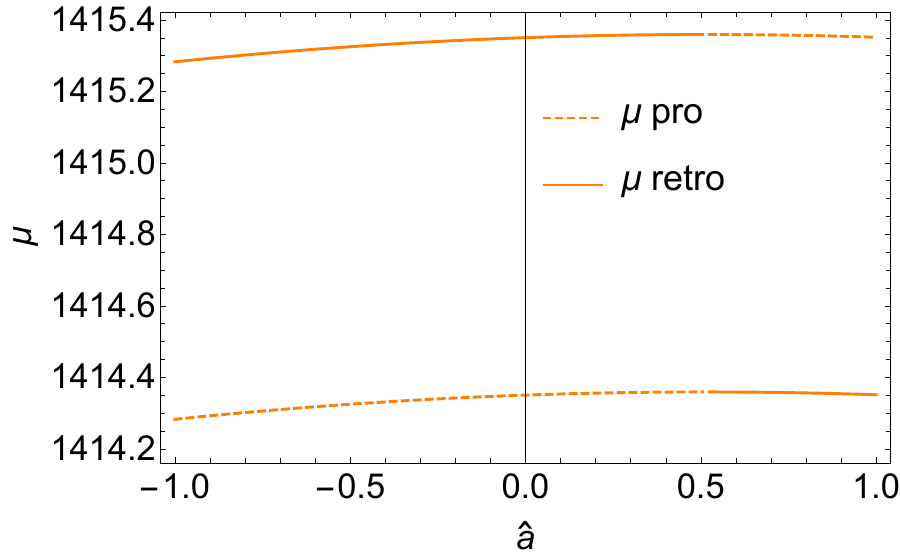}\\
	(c)
	\caption{The dependence of the magnification on $\delta\phi$ and $\delta\theta$ (a) and (b), and on $\hat{a}$ (c). We fixed $\hat{a}=1/2$ in plots (a) and (b) and $\delta\phi=5\times 10^{-6\prime\prime},\,\delta\theta=10^{-3\prime\prime}$ in (c). Other parameters are $\theta_\mas=\pi/4,\,r_\mas=r_\mad,\,v=1$.
	}
	\label{fig:magonparas}
\end{figure}

In Fig. \ref{fig:magonparas} we show the dependence of the magnification on the parameters $\delta\phi,\,\delta\theta$. 
It is seen from Fig. \ref{fig:magonparas} (a) 
that the magnifications $\mu_\pm$ for both images decrease as any of  $\delta\phi$ and $\delta\theta$ increases. Moreover, the magnification $\mu_+$ of the prograde image decreases to zero, while $\mu_-$ of the retrograde image decreases to $1$, which are the asymptotic values of Eq. \eqref{eq:schmag}. 
From Fig. \ref{fig:magonparas} (b), one observes that when both $\delta\phi$ and $\delta\theta$ are small, both magnifications $\mu_\pm$ become very large, a feature qualitatively similar to the magnifications of lensed images in SSS spacetimes but with the total deflection angle $\delta\eta$ in \eqref{eq:totaldef} playing the role of the deflection in SSS spacetime \cite{Xu:2021rld}. However, we see from the peak in Fig. \ref{fig:magonparas} that the value of $(\delta\theta,\,\delta\phi)$ for which the magnifications diverge fastest, do not occur when $\delta\theta\to 0$. This can be attributed to the effect of the spacetime spin. 

In Fig. \ref{fig:magonparas} (c) the effect of $\hat{a}$ on magnification is shown. A small $\delta\phi$ is chosen so that $\hat{a}$'s effect is apparent. In general, it is found that for positive $\delta\phi$, $\mu_\pm$ increases with the increase of $\hat{a}$ up to approximately the critical values $\hat{a}_{c\pm}$. After that, the magnification decreases. The fact that the magnification around $\hat{a}_{c\pm}$ is maximal for each trajectory can be qualitatively understood from the fact that around this spin the trajectory is usually closer to the lens than other $\hat{a}$.

\section{Time delay between images}

To find the time delay between the two lensed images, it is necessary to first compute the total travel time $\Delta t$ along the two trajectories. 
This can be done completely in parallel to the computation of the deflection angle $\Delta\phi$ in Secs. \ref{sec:Preliminaries} and \ref{sec:pertmeth}. Since the computation and presentation there are already quite lengthy, we will address the total travel time and time delay separately here. 

Starting from Eq. \eqref{eq:tgeodesic}, and using Eqs. \eqref{eq:rgeodesic} and \eqref{eq:difftheta} respectively in the first and last terms in the right-hand side of this equation, it becomes 
\begin{align}
	\dd t=&\frac{E(r^2+a^2)^2-2aLMr}{\Delta}\frac{s_{r}\dd r}{\sqrt{R(r)}}  \nn\\
	&-Ea^2 (1-\cos^2\theta) \frac{s_{\theta}\dd \cos\theta}{\sqrt{\Theta(\cos\theta)}} .\label{eq:difft}
\end{align}
The procedure to compute the perturbative $\Delta t$ is then the same as that from Eq. \eqref{eq:diffphi} to Eq. \eqref{eq:Inteddiffphi2} for $\Delta \phi$.
The result is 
\begin{align}
	\Delta t =& \sum_{j=s,d} \sum_{i = -1}^{\infty} H_{r,i}( p_j)\left(\frac{M}{r_0}\right)^i \nn\\
	&+ \sum_{i = 1}^{\infty} H_{\theta,i}^\prime(c_\mas,c_\te) \left(\frac{M}{r_0}\right)^i ,\label{eq:Inteddifft2}
\end{align}
where $H_{r,i}$ and $H^{\prime}_{\theta,i}$ are analogous to the $G_{r,i}$ and $G^{\prime}_{\theta,i}$ in Eq. \eqref{eq:Gr2}-\eqref{eq:Gtheta2}. The first few orders of them are 
\begin{subequations}
	\begin{align}
		H_{r,-1}=&\frac{M\sqrt{1-p_j^2} }{p_j v},\\
		H_{r,0}=&\frac{M}{v^3}\left[\left(3 v^2-1\right) \tanh^{-1}\left(\sqrt{1-p_j^2}\right)+\frac{\sqrt{1-p_j^2}}{p_j+1}\right],\\
		H_{r,1}=& \frac{M \left[15-a^2 \left(c_e^2-2\right)\right] \cos ^{-1}(p_j)}{2 v}\nn\\
		&+\frac{M
			\sqrt{1-p_j}}{2 (p_j+1)^{3/2} v^5} \left\{-6
		(p_j+1) v^2+p_j+2\right.\nn\\
		&\left.-4 s_ls_e  a (p_j+1) v^3 \left[(p_j+1) v^2+1\right]\right\},\\
		H_{\theta,1}^\prime=&\frac{a^2 }{2 v}\left\{-s_{r\theta} \left(c_e^2-2\right) \left[\sin
		^{-1}\left(\frac{c_\mas}{c_e}\right)-a_1\right]\right.\nn\\
		&-\frac{1}{2}s_{r\theta} 
		c_e^2 \sin (2 a_1)+s_{r\theta} c_\mas 
		\sqrt{c_e^2-c_\mas^2}\nn\\
		&\left.+\pi  \left(c_e^2-2\right)
		\left(1-s_{r\theta}\right)\right\},
	\end{align}
\end{subequations}
where $p_j,\, a_1$ are given in Eqs. \eqref{eq:auxdef} and \eqref{eq:a1def} respectively. 

The null limit of $\Delta t$ can be obtained easily by taking $v=1$ in the above equation
\begin{align}
	&\Delta t(v\to 1)=\sum_{j=s,d}\left\{\frac{\sqrt{1-p_j^2}r_0}{p_j}\right.\nn\\
	&+M\left[2 \tanh^{-1}\left(\sqrt{1-p_j^2}\right)+\frac{\sqrt{1-p_j^2}}{p_j+1}\right]\nn\\
	&+\left[\frac{M}{2}\left[15-\hat{a}^2 \left(c_e^2-2\right)\right] \cos ^{-1}(p_j)\right.\nn\\
	&\left.\left.-\frac{M\sqrt{1-p_j}}{2 (p_j+1)^{3/2}} \left[5p_j+4+4 s_l \hat{a} s_e(p_j+1)  \left(p_j+2\right)\right]\right]\left(\frac{M}{r_0}\right)\right\}\nn\\
	&+\frac{\hat{a}^2 }{2}\left\{-s_{r\theta} \left(c_e^2-2\right) \left[\sin
	^{-1}\left(\frac{c_\mas}{c_e}\right)-a_1\right]\right.\nn\\
	&- \frac{s_{r\theta}}{2}
	c_e^2 \sin (2 a_1)+s_{r\theta} c_\mas 
	\sqrt{c_e^2-c_\mas^2}\nn\\
	&+\pi\left(c_e^2-2\right)
	\left(1-s_{r\theta}\right)\Big\}\left(\frac{M}{r_0}\right).
\end{align} 
We have also checked that the $\Delta t$ in Eq. \eqref{eq:Inteddifft2} can reduce to its equatorial plane form computed previously Eq.(53) of Ref. \cite{Liu:2020mkf} if we let $\theta_\mas\to\pi/2,\,\theta_\te\to\pi/2$.

The $\Delta t$ above can be further expanded in the small $p_\mathrm{s,d}$ limit. The result to the first few orders is found to be  
\begin{align}
	\Delta t=&\sum_{n,m_1=-1,m_2=m_1}^{n+m_1+m_2=2}\kappa_{n,m_1,m_2}\lb \frac{M}{r_0}\rb^n \nn\\
	&\times (p_\mas^{m1}p_\mad^{m2}+p_\mad^{m1}p_\mas^{m2})+\mathcal{O}\lb \epsilon^3\rb , \label{eq:dtsmallp}
\end{align}
where the coefficients are
\begin{subequations}
	\begin{align}
		&\kappa_{-1,-1,0}=\frac{M}{v}\label{eq:kappa-1-10},\\
		&\kappa_{-1,0,1}=-\frac{M}{2v}\label{eq:kappa-101},\\
		&\kappa_{0,0,0}=\frac{M}{2v^3}\left\{2+v^2 \ln (64)+\sum_{i=s,d}\left[\ln \left(\frac{p_i }{2}\right)-3 v^2 \ln (p_i )\right]\right\},\label{eq:kappa000}\\
		&\kappa_{0,0,1}=-\frac{M}{v^3},\label{eq:kappa001}\\
		&\kappa_{0,0,2}=-\frac{3M\left(-1+v^2\right)}{4 v^3},\label{eq:kappa002}\\
		&\kappa_{1,0,0}=\frac{M}{4 v^5} \left\{ \left[15 \pi  v^2-12-8 s_l \hat{a} s_e v\left(1+v^2\right) \right]v^2+4\right\},\label{eq:kappa100}\\
		&\kappa_{1,0,1}=\frac{M }{2 v^5}\left\{\left[\left(4 s_l \hat{a} s_e-15 v+\hat{a}^2 v \left(c_e^2-2 c_\mas^2\right)\right)v+6\right]v^2-3\right\},\\
		&\kappa_{2,0,0}=\frac{M}{4 v^7}\left\{-2+6 v^2+v^4 \left[46-15 \pi +70v^2\right.\right.\nn\\
		&+2\hat{a}^2 \left(2+2c_\mas^2-3c_e^2\left(1+v^2\right)-4c_e^2v^2\right)\nn\\
		&\left.\left.-2 s_l s_e \hat{a} v \left(-16+3 \pi\left(4+v^2\right)\right)\right]\right\}.
		\label{eq:tau20res}
	\end{align}
\end{subequations}    
This $\Delta t$ is a function of $r_0$ and $\theta_e$. Therefore once these two quantities, as well as other parameters determining them, are known for a given trajectory, the corresponding $\Delta t$ will be fixed. For the two images formed from the same source but with different $r_{0\pm}$ and $\theta_{\te\pm}$,  the time delay $\Delta^2 t_\pm\equiv \Delta t_+-\Delta t_-$ can be derived through straightforward deduction. 
Using Eq. \eqref{eq:dtsmallp}, 
$\Delta^2 t_\pm$ to the leading three orders is found to be 
\begin{align}
	\Delta^2 t_\pm=&
	-\frac{\left({r_{0+}}^2-{r_{0-}}^2\right) ({r_\mad}+{r_\mas})}{2 {r_\mad} {r_\mas} v}\nn\\
	&+\frac{2 M \left(1-3 v^2\right) }{v^3} \ln\left(\frac{{r_{0+}}}{{r_{0-}}}\right)\nn\\
	&-\frac{M^2}{2  {r_{0+}} {r_{0-}} v^5} \left[\left(15 \pi  v^4-12 v^2+4\right) ({r_{0+}}-{r_{0-}})\right.\nn\\
	&\left.-8 \hat{a} \left(v^2+1\right) v^3 ({r_{0+}}  s_{e-}+{r_{0-}} s_{e+})\right]\nn\\
	&-\frac{M ({r_{0+}}-{r_{0-}}) ({r_\mad}+{r_\mas})}{{r_\mad} {r_\mas} v^3}.
	\label{eq:timedelay}
\end{align}
It is seen that the dominant term (Eqs. \eqref{eq:kappa-1-10}) in Eq. \eqref{eq:dtsmallp} actually does not contribute to the time delay because it is the time corresponds to the straight line approximation and is the same for both trajectories. The terms retained in this analysis originate from Eqs. \eqref{eq:kappa-101}, \eqref{eq:kappa000}, \eqref{eq:kappa100} and \eqref{eq:kappa001}. The effects of spacetime spin and the non-equatorial effects are present in the terms from Eq. \eqref{eq:kappa100}.

\begin{figure}[htp!]
	\centering
	\includegraphics[width=0.45\textwidth]{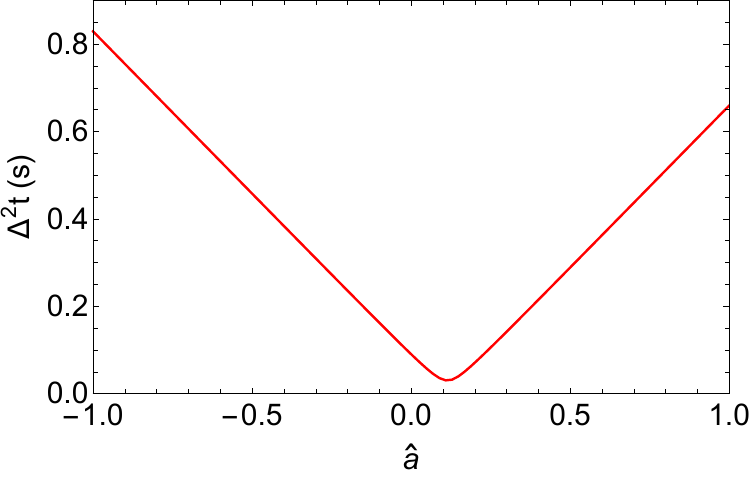}\\
	\caption{The dependence of the time delay $\Delta^2 t_\pm$ on $\hat{a}$ for  $\delta\phi=4^{\prime\prime}~,\delta\theta=1^{\prime\prime},\,r_\mas=r_{s_{39}}=5672.6511(M),\,\theta_\mas=\pi/4,\,v=1$. }
	\label{fig:tdona}
\end{figure}

In Fig. \ref{fig:tdona} we plot the dependence of time delay $\Delta^2 t_\pm$ on $\hat{a}$ for a source with  $\delta\phi=4^{\prime\prime}$ and $\delta\theta=1^{\prime\prime}$, and the lens is still assumed to be Sgr A*. 
Since $\delta\phi$ is very small, as indicated in Fig. \ref{fig:r0thetamonhata} and known in Ref. \cite{Liu:2020mkf}, the spin $\hat{a}$ is expected to have a significant impact on the time delay. 
It is seen then as $\hat{a}$ varies from $-1$ to 1, the time delay changes from approximately $0.025$ (s) to $0.83$ (s). 
The time delay reaches its minimum value around $\hat{a} = 0.11$, which is close to the critical value $\hat{a}_{c\pm}$. In other regions, the time delay varies with roughly a constant size but appreciable slope. 

\section{Applications\label{sec:appl}}

In this section, we discuss a few problems that our results can be used to study. 

\subsection{Image tracks of a moving source}

We will first study the images of a source moving in the equatorial plane behind the lens. Such sources can include stars or other transits whose orbit (almost) intersect with the observer-lens axis and whose angular velocity is appreciable so that observation of the motion becomes possible, e.g., some S stars around Sgr A*. 

\begin{figure}[htp!]
	\centering
 \includegraphics[width=0.4\textwidth]
	{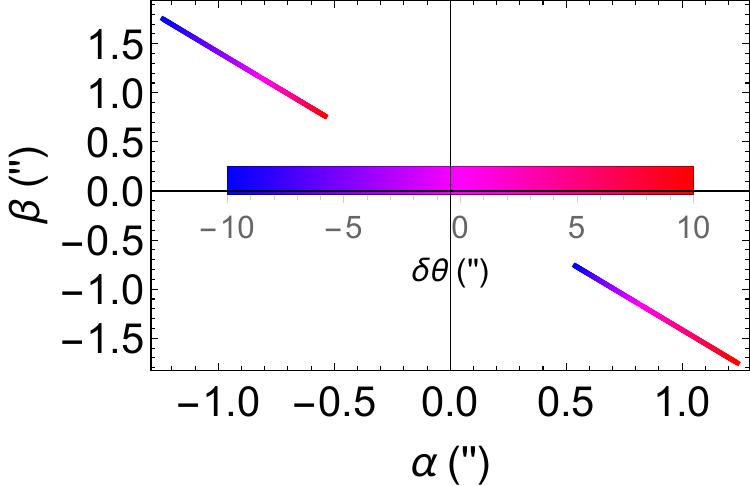}\\
	(a)\\
\includegraphics[width=0.4\textwidth]
	{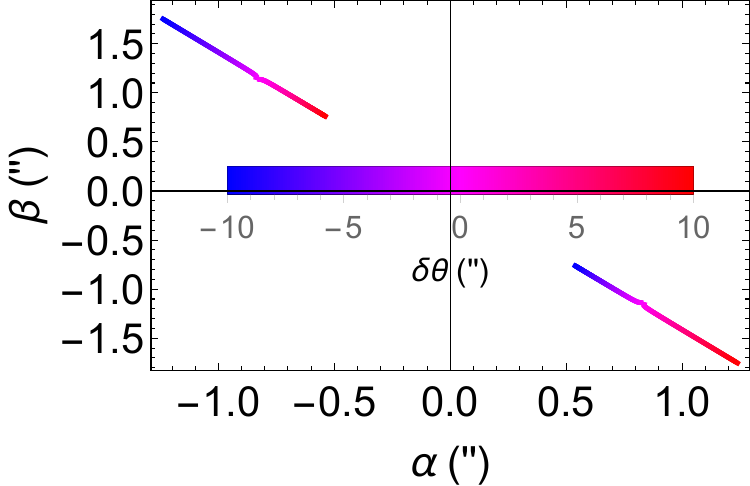}\\
	(b)\\
\includegraphics[width=0.4\textwidth]
	{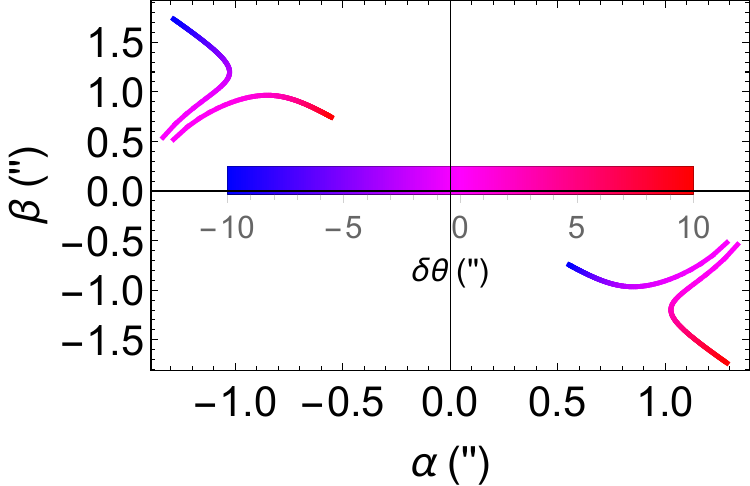}\\
 (c)
	\caption{The dependence of the apparent image and magnification on moving sources, where $\delta\phi=\delta_0+\delta\theta$ and $\delta\theta$ from $-10^{\prime\prime}$ to $10^{\prime\prime}$. We fixed $\hat{a}=1/2,\theta_\mas=\frac{\pi}{4},\,r_\mas=r_\mad,\,v=1$ and $\delta_0=10^{-5\prime\prime}$ (a), $\delta_0=10^{-3\prime\prime}$ (b), $\delta_0=10^{-1\prime\prime}$ (c) in this figure.}
	\label{fig:sstarmotion}
\end{figure}

In Fig. \ref{fig:sstarmotion}, we assume that the source is located at some representative radial distance of the S star orbits, and plot the location of the GL image of this source as it moves across. Since we are working within a WDL, basically the section of the trajectory that we can treat looks almost a straight line if there was no lens. We assume that this straight line satisfies the parametric relation $\delta\phi=\delta_0+\delta\theta$ where $\delta_0$ takes on a few values $10^{-5\prime\prime},10^{-3\prime\prime},10^{-1\prime\prime}$ and $\delta\theta$ runs from $-10^{\prime\prime}$ to $10^{\prime\prime}$ in Fig. \ref{fig:sstarmotion} in order to compute the corresponding image locations. 

It is seen that when $\delta_0$ is relatively small, i.e for the $\delta_0=10^{-3\prime\prime},\,10^{-5\prime\prime}$ cases, the images of a source moving along a straight line in the backend also form two straight tracks on the celestial sphere. This indicates that in this parameter setting, the effect of the spacetime spin is not important in determining the apparent angles of the images. When $\delta_0$ is larger (i.e., $10^{-1^\prime\prime}$) however, the image tracks deviate from straight lines at some point of $\delta\theta$ or $\delta\phi$. This is indeed the value of $\delta\theta$ and $\delta\phi$ such that the critical $a_c=1/2$, the value we set for the spacetime spin when plotting this figure. This sharp derivation from straight lines of the tracks can be used as a characteristic observable of $a_c$.

\subsection{Shape of lensed images}

Among possible sources lensed by a Kerr BH, stars or other spherical shape objects are very natural candidates. If the source and/or the lens are small and too far from us so that the shape of the images are not resolvable, then only the central values of the apparent angles $(\alpha,\,\beta)$ might be obtained. 
However when the source is large or close and the detector has enough resolution, then the shape of the source should also be recognizable. In an SSS spacetime, then one would expect that the images of a spherical source will generally appear elongated. 

\begin{figure}[htp!]
	\centering
	\includegraphics[width=0.4\textwidth]
	{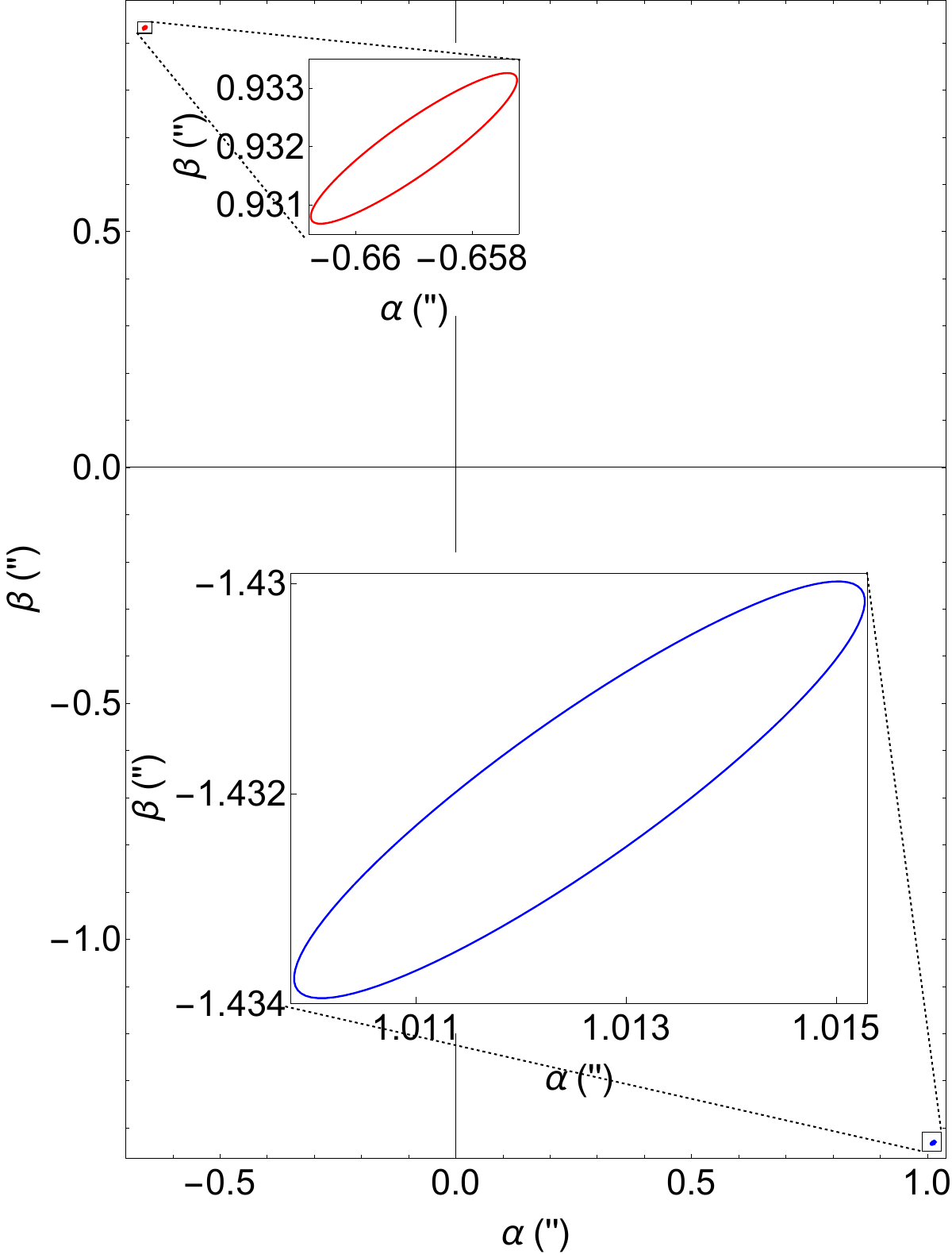}
	\caption{Apparent angles $(\alpha,\,\beta)$ of the two images of a star with twice the Sun's size. We fixed $s_{r\theta}=+1,\,s_{l}=+1,\,\theta_\mas=\pi/4,\,\delta\phi=\delta\theta=1^{\prime\prime} ,\,\hat{a}=1/2,\,r_\mas=r_\mad,\,v=1$. The red, blue and black dashed lines are the two lensed images and the source shape without the presence of the lens (the inset is the zoomed-in images).}
	\label{fig:shapeimage}
\end{figure}

In the Kerr spacetime, if the $\delta\theta$ and $\delta\phi$ are large so that the effect of $\hat{a}$ is much weaker on the trajectory deflection, then naturally one would expect that the shape of the images will be similar to those in the Schwarzschild spacetime with $\delta\eta$ in Eq. \eqref{eq:totaldef} playing the rule of the source's deflection.
In this case, we can work out the shape of the images based on the apparent angle formula \eqref{eq:schimg} in the Schwarzschild spacetime. 
Supposing that the source object's radial coordinate is $R$ where $R/r_\mas\ll \delta\eta$ (see Fig. \ref{fig:shapeimage}), and denoting the polar angle of a boundary point of the source to be $\sigma$ in a polar frame with the center of the source at the origin,  then to this leading order of $R/r_\mas$, the total deflection angle of this boundary point with respect to the Schwarzschild lens becomes 
\begin{align}
	\delta\eta^\prime =\delta\eta+\frac{R}{r_\mas}\sin\sigma+\mathcal{O}\lb \frac{R}{r_\mas}\rb^2,~\sigma\in[0,2\pi). 
\end{align}
Substituting this into Eq. \eqref{eq:schimg} and to the first order of $R/r_\mas$, the boundaries of the two images now have apparent angles $(\alpha_{S,\pm}^\prime,\,\beta_{S,\pm}^\prime)$ with 
\begin{align}
	\alpha_{S,\pm}^\prime=&\alpha_{S,\pm}+\frac{R}{2(r_\mad+r_\mas)}
	\lb\text{sgn}(\delta\phi)\mp \frac{1}{\zeta} \rb\sin\sigma\\
	\equiv & \alpha_{S,\pm}+\delta\alpha_{\pm},~\sigma\in[0,2\pi)\label{eq:alphavar}
\end{align}
where $\alpha_{S,\pm}$ is in Eq. \eqref{eq:schimg} and $\zeta$ in Eq. \eqref{eq:schimg2}. The last term is the small variation $\delta\alpha_{\pm}$ of the apparent angle in the $\alpha$ direction of the images. Similarly, the extension of the images in the $\beta$ direction is 
\begin{align}
	\delta\beta_\pm=\frac{R\cos\sigma}{r_\mas}\frac{\alpha_{S,\pm}}{\delta\eta}.\label{eq:betavar}
\end{align}

Eqs. \eqref{eq:alphavar} and \eqref{eq:betavar} clearly demonstrate that the images on the celestial sphere have elliptic shapes with $\delta\alpha_{\pm}$ and $\delta\beta_{\pm}$ being the semi-minor and semi-major axes respectively, with the semi-minor axis aligned with the image-lens axis. The eccentricity of these ellipses are
\begin{align}
	e_\pm=\frac{|\delta\beta_{\pm}/\cos\sigma|-|\delta\alpha_{\pm}/\sin\sigma|}{|\delta\beta_{\pm}/\cos\sigma|+|\delta\alpha_{S,\pm}/\sin\sigma|}=\frac{\zeta-1}{\zeta+1} .\label{eq:eres}
\end{align}
The corresponding magnifications of the images are the ratio between the angular size of the images and the original source 
\begin{align}
	\mu_\pm=&\left|\frac{\pi\delta\alpha_{S,\pm}/\sin\sigma \cdot \delta\beta_\pm/\cos\sigma}{ \pi R^2/(r_\mas+r_\mad)^2}\right|\nn\\
	&=\pm\frac14 \lb 1\pm \text{sgn}(\delta\phi)\frac{1}{\zeta}\rb\lb 1\pm \text{sgn}(\delta\phi)\zeta\rb. \label{eq:mures}
\end{align}
One can check that this agrees with the magnification  \eqref{eq:schmag} for images of the source at $\delta\eta$.

In Fig. \ref{fig:shapeimage} we show the image locations of a star with twice the size of our Sun and located at $r_\mas=100^{\prime\prime}r_\mad$ and $\delta\theta=\delta\phi=1^{\prime\prime}$. One can see that they do take the elliptic shape and we have checked that their eccentricity and magnifications match exactly the values specified by Eqs. \eqref{eq:eres} and \eqref{eq:mures}.

\subsection{Constraining the BH orientation}

Kerr BH spacetime is thought as the most important BH in astronomy, while the SMBH Sgr A* is currently one of the best confirmed BH candidates. Even being so close to us, Sgr A* still has many properties not well-constrained, including its spin orientation against our line-of-sight. 

However, if the images of a source that is well aligned with the detector-lens axis are observed, then we might attempt to constrain the inclination of the spin, which is given by $\theta_i=\pi/2-\theta_s$ in this case. Among the input parameters $\{M,\,a,\,\theta_\mas,\,r_\mas,\,r_\mad,\,\delta\theta,\,\delta\phi, ~v\}$, the parameters $M,\,\hat{a}$ and $\theta_\mas$ are associated with the BH itself. The parameters $r_\mas,\,\delta\theta$ and $\delta\phi$ are associated with the source and $r_\mad,\,v$ are associated with the detector andtest particles respectively. Generally, the parameters $M, ~r_\mad$ can be obtained through other means and we can set $v=1$ for photons. $\delta\theta,\,\delta\phi$ normally can not be measured while $r_\mas$ can sometimes be deduced from the spectrum redshift if the source is a far-away galaxy but would be more difficult to measure for a typical star in the Galaxy. The SgrA* spin has been measured although not tightly \cite{Fragione:2022oau} and therefore we mainly try to constrain its orientation against the line-of-slight in this paper.

We assume that for a GL situation by a Kerr BH, we can observe either the angle  $\sigma$ between the line connecting the two images and the projection of $\hata$ on the celestial sphere, i.e., 
\begin{align}
    \sigma=\arctan (\beta_\pm/\alpha_\pm),
\end{align}
or the time delay $\Delta^2 t_\pm$ between the two images. Each of these two quantities then will allow us to solve $\theta_s$ and consequently the inclination $\theta_i$. We have listed a few typical values of the observed $\sigma$ or $\Delta^2t_\pm$ and the deduced  $\theta_i$ in Tab. \ref{tb:solve}. It is seen that $\theta_i$ depend on $\sigma $ and $\Delta^2 t_\pm$ very sensitively and therefore can be well constrained by them.  

\begin{table}[]
\centering
\begin{tabular}{p{1.5cm}|p{1.5cm}||p{1.5cm}|p{1.5cm}}
			\hline
			\hline
 $\Delta^2 t_\pm$ (s)&$\theta_i$ (rad) & 			$\sigma 
 $ (rad) & $\theta_i$ (rad)\\
			\hline
		0.2  &1.31&0.05 &0.67\\
   \hline
   		0.3  &1.17&0.10 &1.17\\
     \hline
   		0.4  &1.02&0.15 &1.31\\
     \hline
   		0.5  &0.86&0.20 &1.38\\
     \hline
   		0.6  &0.67&0.25 &1.42\\
   		\hline
			\hline
\end{tabular} 
\caption{Deduction of BH spin inclination $\theta_i=\pi/2-\theta_s$ from $\Delta^2t$ or $\sigma$. Columns 1 and 3 are assumed measurements and columns 2 and 4 are solved $\theta_i$. We fix $\delta\theta=1^{\prime\prime},\,\delta\phi=10^{-5\prime\prime} ,\,r_\mas=r_{s39},\,v=1,\,\hata=0.7$}
\label{tb:solve}
\end{table}
 
\section{Conclusion and Discussions}

This work considered the deflections and GL of both null signals and massive particles in the off-equatorial plane in Kerr spacetime in the WDL. The deflection angles are computed using the perturbative method resulting in power series expansions of $M/r_0$ and $r_0/r_{\mathrm{s,d}}$ with the coefficients being functions of the spacetime parameter and the trigonometric function of the extreme values $\theta_\te$ along the trajectories. Also importantly, the finite distance effect of the source and detector is taken into account in these deflections. This allows us to establish a set of exact GL equations, from which we can solve the desired $(r_0,\theta_\te)$ allowing the test particle from a source with deviation angles $\delta\theta$ and $\delta\phi$ to reach the detector. 

Using the exact formula for the apparent angles derived for the off-equatorial plane test particles, we studied the effect of various parameters, including the source deviation angles $\delta\theta$ and $\delta\phi$, the spacetime spin $\hat{a}$ and its orientation $\theta_\mas$, on the angular locations of the images on the celestial sphere and their magnifications. It is generally found that there exist usually two trajectories that will reach the detector. For given values of $\delta\theta$ and $\delta\phi$, there always exist two critical values of $\hat{a}_c$ at which the two test particles intersect the positive and negative $\hat{z}$ directions respectively. Generally, when $\delta\theta$ or $\delta\phi$ are large, the effect of $\hat{a}$ (for the Kerr BH $\hat{a}\leq1$) becomes subdominant and therefore the GL is roughly the same as that in the Schwarzschild spacetime (but in the off-equatorial plane). However in other cases, $\hat{a}$ will affect the quadrant in which the images appear, and the magnification of these images. 

We also obtained the time delays between the two images. It was found that in general, the time delay depends on the spacetime spin $a$ very sensitively even when the deviation angles $\delta\theta$ and $\delta\phi$ are not very small. Therefore it can be used as a good tool to constrain $a$ as was shown in the equatorial case \cite{Liu:2020mkf}. 

We used these results to study the image of a transiting source behind the lens, the image shape and size of a spherical source and used the observables to deduce properties of the BH and its orientation. 

There are a few points that we can discuss here. The first is that although in this work we mainly studied the black hole spacetime with $|a|\leq M$, it is actually not restricted to this range. Therefore the method and results are also applicable to the naked singularity case. 
The second point is that, from the mathematical point of view, the perturbation method should be generalizable to the deflection and GL in the off-equatorial plane of other axisymmetric spacetime. We will report the findings along this direction in a follow-up work. 

\acknowledgements

The work of T. Jiang and X. Xu is partially supported by the Wuhan University Students Innovation and Entrepreneurship Program.

\appendix

\section{Integrability of the series \label{sec:appd}}

In this appendix, we show that the integrals of the forms \eqref{eq:pintgform} and \eqref{eq:cintgform} can always be carried out and the results are elementary functions. 

For the integral \eqref{eq:pintgform}, multiplying the numerator and denominator of the integrand by $(1-p)^{i-1}$, the integral is transformed to the sum of integrals of the form of the left-hand side of the following equation
\begin{align}
	& \int_1^{p_\mathrm{s,d}} \frac{p^k\dd p}{\left(1-p^2\right)^{i-1/2}}=\left.-\frac{p^{k-1}}{(2i+k-4)\left(1-p^2\right)^{i-3/2}}\right|_1^{p_\mathrm{s,d}}\nonumber\\
	+& \frac{k-1}{2i+k-4}\int_1^{p_\mathrm{s,d}}\frac{p^{k-2}\dd p}{\left(1-p^2\right)^{i-1/2}},~~(k+1,i=1,2,\cdots)\label{eq:recrel1}
\end{align}
where the integration is done by parts. Note that superficially the first term on the right-hand side might diverge when $p$ approaches $1$. However, all these divergences will cancel when substituting these results into Eq. \eqref{eq:pintgform} because this is an artifact introduced when multiplying its integrand denominator by $(1-p)^{i-1}$. The recursion relation \eqref{eq:recrel1} allows us to lower the order of the numerator by 2. Finally, for the lowest two orders $k=0,\,1$ cases, we have
\begin{align}
	\int_1^{p_\mathrm{s,d}} \frac{\dd p}{\left(1-p^2\right)^{i-1/2}} =&\sum^{i-2}_{k=0}\frac{C^{k}_{i-2}}{2k+1} \left.\frac{p^{2k+1}}{\left(1-p^2\right)^{i+1/2}}\right|_1^{p_\mathrm{s,d}},\\
	\int_1^{p_\mathrm{s,d}} \frac{p\dd p}{\left(1-p^2\right)^{i-1/2}} =&\left.-\frac{1}{(2i-3)\left(1-p^2\right)^{i-3/2}} \right|_1^{p_\mathrm{s,d}}.
\end{align}
From the above relations, we see that the result of the integration \eqref{eq:pintgform} is a sum of terms like $p^m/(1-p^2)^{n-1/2}~(m,n=1,2,\cdots)$, which are elementary functions.

For the integral \eqref{eq:cintgform}, using a further change of variable $x=c/c_\te$, it becomes after integration by parts and simplification
\begin{align}
	&\int_{c_\mathrm{s,d}/c_\te}^1 \frac{x^{2 n}}{\sqrt{1-x^2}}\dd x = \left.-\frac{x^{2n-1}\sqrt{1-x^2}}{2n}\right|_{c_\mathrm{s,d}/c_\te}^1\nonumber\\
	&+\frac{2n-1}{2n}\int_{c_\mathrm{s,d}/c_\te}^1 \frac{x^{2 n-2}}{\sqrt{1-x^2}}\dd x,~(n=1,2,\cdots).
\end{align}
Using this recursion relation and the lowest order integral
\begin{align}
	\int_{c_\mathrm{s,d}/c_\te}^1 \frac{1}{\sqrt{1-x^2}}\dd x = &\left.\sin^{-1}{x}\right|_{c_\mathrm{s,d}/c_\te}^1,
\end{align}
we see that \eqref{eq:cintgform} can also be completely integrated and the result is a sum of elementary functions. 

\section{Derivation of $\cos(\theta_\mad)$ in Eq. \eqref{eq:cdfinal} 
	\label{appd:cdr0rel}}

In this appendix, we present two approaches for the derivation of $\cos(\theta_\mad)$ in terms of other kinematic parameters. The resultant Eq. \eqref{eq:cdfinal} will act as one of the two GL equations from which the $\theta_\te$ and $r_0$ can be solved. The first approach is the Jacobian elliptic function method, which directly solves the Eq. \eqref{eq:thetaintdef} for $\cos\theta_\mad$. The second approach uses the method of undetermined coefficients.

For the first method, inspecting Eqs. \eqref{eq:Rdef} and \eqref{eq:Thetadef} and focusing on their dependence on $c$ and $r$ respectively, we can factor them as
\begin{align}
	\Theta(c)=&(c^2_m-c^2)(B_0c^2+B_1),\\
	R(r)=&(E^2-m^2)\left[(r-r_0)(r-r_1)(r-r_2)(r-r_3)\right],
\end{align}
where $c_\te$ and $r_0,\,r_1,\,r_2,\,r_3$ are roots of $\Theta(c)=0$ and $R(r)=0$ respectively and we ordered them as $r_0>r_1>r_2>r_3$. The coefficients $B_0$ and $B_1$ are 
\begin{align}
	&B_0=a^2 v^2 E^2,\nn\\
	&B_1=v^2r_0^2 E^2+\frac{2 M r_0 E^2\left[\Sigma(r_0,\theta_\te)+a^2s_\te^2(1+v^2)\right]}{\Sigma(r_0,\theta_\te)-2Mr_0}\nn\\
	&~~~~~~+\frac{4 a M s_\te r_0 E^2}{(\Sigma(r_0,\theta_\te)-2 M r_0)^2}\left(2 a M r_0 s_\te\right.\nn\\
	&\left.-s_l\sqrt{\Delta(r_0)\Sigma(r_0,\theta_\te)\left[(\Sigma(r_0,\theta_\te)-2Mr_0)v^2+2 M r_0\right]}\right).\nn
\end{align}
With this re-writing, then Eq. \eqref{eq:difftheta} can be solved to find a solution of $\cos(\theta)$ as a function of $r$ 
\begin{align}
	\cos(\theta)=c_\te\,\mathrm{cn}\left(F(r)+C\left|\frac{B_0c^2_m}{B_0 c^2_m+B_1}\right.\right),
	\label{eq:diffjacobi-general}
\end{align}
where $\mathrm{cn}(x|y)$ is the Jacobian elliptic function, $C$ is the integral constant and
\begin{align}
	&F(r)=\frac{2(B_0 c^2_m+B_1)}{(E^2-m^2)\sqrt{(r_0-r_2) (r_1-r_3)}} \nn\\
	\times&  F_1\left[\sin ^{-1}\left(\sqrt{\frac{(r_1-r_3) (r-r_0)}{(r_0-r_3)
			(r-r_1)}}\right)|\frac{(r_1-r_2) (r_0-r_3)}{(r_0-r_2)
		(r_1-r_3)}\right]\nn,
\end{align}
where $F_1(x|y)$ is the elliptic integral of the first kind. 

To fix the constant $C$, we use the boundary condition $\theta(r=r_\mas)=\theta_\mas$ to find
\begin{align}
	C=\mp\left[F (r_\mas)- s_{r,\theta}\mathrm{cn}^{-1}\left(\frac{c_\mas}{c_\te}
	\left|\frac{B_0c^2_m}{B_0 c^2_m+B_1}\right. \right)\right]\label{eq:intcfixed}
\end{align}
where the $-$ and $+$ signs in $\mp$ correspond to the branch of trajectory from $r_\mas$ to $r_0$ and from $r_0$ to $r_\mad$ respectively.
Substituting Eq. \eqref{eq:intcfixed} and $r=r_\mad$ into solution \eqref{eq:diffjacobi-general} and expanding the result in terms of small $M/r_0$, we then can obtain the series \eqref{eq:cdfinal}. 

In the second method to determine the relation between $c_\mad$ and $r_0$,  we start by assuming that $c_\mad$ takes a series form as in Eq. \eqref{eq:cdfinal} and then use the method of undetermined coefficients to determine these $h_i$. 

Substituting this series form into the right-hand side of Eq. \eqref{eq:Inteddifftheta} and then recollecting the series according to the power of $(M/r_0)$, one obtains
\begin{align}
	\sum_{i=1}^{\infty} F_{r,i}(p_\mas,p_\mad) \left(\frac{M}{r_0}\right)^i &= \sum_{i=1}^{\infty} F_{\theta,i}(c_\mas,c_\mad,c_\te) \left(\frac{M}{r_0}\right)^i \nn\\
	&= \sum_{i=1}^{\infty} F_{\theta,i}^\prime(c_\mas,c_\te) \left(\frac{M}{r_0}\right)^i,
\end{align}
where the first few $F_{\theta,i}^\prime$ are listed as
\begin{align}
	F_{\theta,1}^\prime =& \frac{1}{\sqrt{E
			^2-m^2}}\left[\tan ^{-1}\left(\frac{h_0 c_\mas}{\sqrt{c_\te^2-h_0^2 c_\mas^2}}\right)\right.\nn\\
	&\left.  +\sin ^{-1}\left(\frac{c_\mas}{c_\te}\right) s_{r\theta} \pi\right],\\
	F_{\theta,2}^\prime =& \frac{- h_1 c_\mas \left(E^2-m^2\right)}{\left(E ^2-m^2\right)^{3/2} \sqrt{c_\te^2-h_0^2 u_\mas^2}} \nn\\
	& + \frac{E^2}{\left(E ^2-m^2\right)^{3/2}}\left[\tan ^{-1}\left(\frac{h_0 c_\mas}{\sqrt{c_\te^2-h_0^2 c_\mas^2}}\right)\right.\nn\\
	&\left.+s_{r\theta} \sin ^{-1}\left(\frac{c_\mas}{c_\te}\right)  \pi \right].
\end{align}
The coefficients $h_i$ here can be fixed by comparing with the left-hand side of Eq. \eqref{eq:Inteddifftheta} for the coefficient of $(M/r_0)^n$ order by order, i.e.,
\begin{equation}
	F_{r,i}(p_\mas,p_\mad) = F_{\theta,i}^\prime(c_\mas,c_\te) \qquad \text{for $i=1,2,3...$}.  \label{eq:eqgG}
\end{equation}
Fortunately, this set of system can be solved iteratively because $h_i$ always starts to appear from $F_{\theta,i+1}^\prime$ and the equation system is simple enough. The result of the solutions to $h_i$ are exactly Eqs. \eqref{eq:hresulth2}. 

\section{Derivation of the apparent angles\label{sec:appdappang}}

First, we denote the four-velocity of the test particle and a direction between which we want to measure the angle, as $v^\mu$ and $k^\mu$ respectively, then we have
\begin{align}
	v^\mu=\lb \dot{t},~\dot{r},~\dot{\theta},~\dot{\phi} \rb.\label{eq:4-p}
\end{align}
For $k^\mu$, there are naturally three choices, the directions $\hat{r}_\mad,\,\hat{\theta}_\mad$ and $\hat{\phi}_\mad$ which are the spacelike directions of the tetrad $e^\mu_{(a)}$ associated with a static observer with four velocity $u^\mu$
\begin{align}
	&e^\mu_{(1)} =\hat{r}_\mad=\lb 0,~\frac{\Delta_\mad}{\Sigma_\mad},~0,~0\rb ,\label{eq:rsobs}\\
	&e^\mu_{(2)} =\hat{\theta}_\mas=\lb 0,~0,~\frac{1}{\Sigma_\mad},~0\rb, \label{eq:thetasobs}\\
	&e^\mu_{(3)} =\hat{\phi}_\mas=\sqrt{\frac{\Sigma_\mad-2Mr_\mad}{\Delta_\mad\Sigma_\mad}}\lb -\frac{2 a M r_\mad s_\mad^2}{\Sigma_\mad-2Mr_\mad},~0,~0,~\frac{1}{s_\mad}\rb, \label{eq:phisobs}\\
	&e^\mu_{(0)}= u^\mu =\lb \frac{\Sigma_\mad}{\Sigma_\mad-2Mr_\mad},~0,~0,~0\rb. \label{eq:4-p-obs} 
\end{align}
For such static observers, 
we can use the projection operators \begin{align}
	P_{\mu\nu}=g_{\mu\nu}+u_{\mu}u_\nu \text{~and~} Q_{\mu\nu}=u_{\mu}u_\nu,
\end{align}
to project each of the test particle or directional vectors $v^\mu,~\hat{r}_\mad,\,\hat{\theta}_\mad,\,\hat{\phi}_\mad$ into a spacial part and a temporal part in the rest frame of the observer respectively \cite{bk:Soffel}
\begin{align}
	&v_\mu= P_{\mu\nu}v^\nu +Q_{\mu\nu}v^\nu,\\
	&k_\mu= P_{\mu\nu}k^\nu +Q_{\mu\nu}k^\nu,~k=\hat{r}_\mad,~\hat{\theta}_\mad,~\hat{\phi}_\mad.
\end{align}
Then the apparent angle of the signal against $\hat{r}_\mad$ is given by
\begin{align}
	\gamma=\cos^{-1} \frac{(\bar{v},\bar{\hat{r}}_\mad)}{|\bar{v}||\bar{\hat{r}}_\mad|},~\text{where}~\bar{X}_\mu=P_{\mu\nu}X^\nu
\end{align}
and the angle between the signal and the $\hat{r}_\mad\hat{\theta}_\mad$ plane and the $\hat{r}_\mad\hat{\phi}_\mad$ plane 
are respectively 
\begin{align}
	&\alpha=\sin^{-1} \frac{(\bar{v},\bar{\hat{\phi}}_\mad)}{|\bar{v}||\bar{\hat{r}}_\mad|},\\
	&\beta=\sin^{-1} \frac{(\bar{v},\bar{\hat{\theta}}_\mad)}{|\bar{v}||\bar{\hat{\theta}}_\mad|}.
\end{align}

Substituting the Kerr metric, we are able to find the expressions for these three angles as
\begin{align}
	\alpha=&\sin^{-1} \frac{L(\Delta_\mad-a^2 s_\mad^2)+2 a M \E  r_\mad s_\mad^2}{s_\mad\sqrt{\Delta_\mad\Sigma_\mad(\E ^2\Sigma_\mad-m^2(\Delta_\mad-a^2 s_\mad^2))}},\label{eq:anglegammaphi}\\
	\beta=&\sin^{-1} \frac{s_{r\theta}\sqrt{\Theta(c_\mad)(\Delta_\mad-a^2 s_\mad^2)}}{s_\mad\sqrt{\Sigma_\mad\lsb \E ^2\Sigma_\mad-m^2(\Delta_\mad-a^2 s_\mad^2)\rsb}},\label{eq:anglegammatheta}\\
	\gamma=&\cos^{-1} \left\{ \frac{[(a L-(a^2+r_\mad^2)\E )^2-(K+m^2 r_\mad^2)\Delta_\mad]}{\Delta_\mad\Sigma_\mad\lsb \E ^2\Sigma_\mad-m^2(\Delta_\mad-a^2 s_\mad^2)\rsb}\right.\nn\\
	&~~~~~~~~\times (\Delta_\mad-a^2s_\mad^2)\bigg\}^{1/2}.\label{eq:anglegammar}
\end{align}

\end{document}